\documentclass{aa}
\usepackage{graphicx}
\usepackage{txfonts}  

\newcommand{\Teff}{T_{\rm eff}}

\newcommand{\Vmic}{V_{\rm mic}}
\newcommand{\Vmac}{V_{\rm mac}}
\newcommand{\eps}[1]{\log\varepsilon_{\rm #1}}
\newcommand{\kms}{km\,s$^{-1}$}
\newcommand{\kH}{$S_{\rm H}$}
\begin{document}

\title{Non-LTE line formation for heavy elements in four very metal-poor stars
\thanks{Based on observations collected at Subaru Telescope, which is operated by the National
Astronomical Observatory of Japan}}

\author{L. Mashonkina \inst{1,2}
\and G. Zhao \inst{3} \and T. Gehren \inst{1} \and W. Aoki
\inst{4} \and M. Bergemann \inst{1} \and K. Noguchi \inst{4} \and
J.R. Shi \inst{3} \and M.Takada-Hidai \inst{5} \and H.W. Zhang
\inst{6} }

\offprints{L. Mashonkina}

\institute{ Institut f\"ur Astronomie und Astrophysik der
Universit\"at
M\"unchen, Scheinerstr. 1, 81679 M\"unchen, Germany\\
\email{lyuda@usm.lmu.de}
\and Institute of Astronomy, Russian Academy of Science, Pyatnitskaya 48,
119017 Moscow, Russia\\
\email{lima@inasan.ru} \and National Astronomical Observatories,
Chinese Academy of Sciences, A20 Datun Road, Chaoyang District,
Beijing 100012, PR China \and National Astronomical Observatory,
Mitaka, Tokyo 181-8588, Japan \and Liberal Arts Education Center,
Tokai University, Hiratsuka, Kanagawa 259-1292, Japan
\and Department of Astronomy, School
of Physics, Peking University, Beijing 100871, PR China}

\date{Received  / Accepted }

\abstract{ {\it Aims.} Stellar parameters and abundances of Na,
Mg, Al, K, Ca, Sr, Ba, and Eu are determined for four very
metal-poor (VMP) stars ($-2.66 \le$ [Fe/H] $\le -2.15$). For two
of them, HD\,84937 and HD\,122563, the fraction of the odd
isotopes of Ba derived for the first time.

\noindent {\it Methods.} Determination of an effective
temperature, surface gravity, and element abundances was based on
non-local thermodynamic equilibrium (non-LTE) line formation and
analysis of high-resolution (R $\sim60000$ and 90000) high
signal-to-noise (S/N $\ge 200$) observed spectra. A model atom for
\ion{H}{i} is presented. An effective temperature was obtained
from the Balmer H$_\alpha$ and H$_\beta$ line wing fits. The
surface gravity was calculated from the {\sc Hipparcos} parallax
if available and the non-LTE ionization balance between
\ion{Ca}{i} and \ion{Ca}{ii}. Based on the hyperfine structure
affecting the \ion{Ba}{ii} resonance line $\lambda\,4554$, the
fractional abundance of the odd isotopes of Ba was derived from a
requirement that Ba abundances from the resonance line and
subordinate lines of \ion{Ba}{ii} must be equal.

\noindent {\it Results.} For each star, non-LTE leads to a
consistency of $\Teff$ from two Balmer lines and to a higher
temperature compared to the LTE case, by up to 60~K. Non-LTE
effects are important in spectroscopic determination of surface
gravity from the ionization balance between \ion{Ca}{i} and
\ion{Ca}{ii}. For each star with a known trigonometric surface
gravity, non-LTE abundances from the lines of two ionization
stages, \ion{Ca}{i} and \ion{Ca}{ii}, agree within the error bars,
while a difference in the LTE abundances consists of 0.23~dex to
0.40~dex for different stars. Departures from LTE are found to be
significant for all investigated atoms, and they strongly depend
on stellar parameters. For HD\,84937, the Eu/Ba ratio is
consistent with the relative solar system $r-$process abundances,
and the fraction of the odd isotopes of Ba, $f_{odd}$, equals
0.43$\pm$0.14. The latter can serve as an observational constraint
on $r-$process models. The lower Eu/Ba ratio and $f_{odd}$ =
0.22$\pm$0.15 found for HD\,122563 suggest that the $s-$process or
the unknown process has contributed significantly to the Ba
abundance in this star.

 }

\authorrunning{L. Mashonkina, et al.}
\titlerunning{Non-LTE line formation for heavy elements in metal-poor stars}

\maketitle

\section{Introduction}

Very metal-poor stars (hereafter VMP stars, with
[Fe/H]\footnote{[A/B] = $\log(N_A/N_B)_* - \log(N_A/N_B)_\odot$
where $N_X$ are number densities.} $< -2$) belonging to the halo
population of the Galaxy were formed at the earliest epoch of star
formation. These stars preserve in the atmosphere the chemical
composition produced by the early generation of stars.
Determination of element abundances in VMP stars is important for
understanding nucleosynthesis processes in the early Galaxy. The
more metal-poor stars are, on average, more distant, and they are
more difficult to analyze than nearby stars with metal abundances
close to solar. In particular, the methods of surface gravity
determination are of limited usefulness. For distant field halo
stars, trigonometric parallaxes are unknown. Progress in this
direction is expected due to the upcoming ESA Gaia satellite
mission (Perryman et al.~2001). The strongest lines of neutral
magnesium (the \ion{Mg}{i}b lines) and neutral calcium (the
\ion{Ca}{i}\, triplet $4p - 5s$) show no wings, and, therefore,
they are not sensitive to surface gravity. In fact, surface
gravities of VMP stars can only be found from the ionization
balance between neutral and ionized species of a selected atom.
One mostly selects iron and Fe peak elements. In two recently
found hyper metal-poor stars (hereafter HMP stars, with [Fe/H] $<
-5$), only Ca is observed in two ionization stages (Christlieb et
al. \cite{chri02}, Frebel et al. \cite{frebel05}) and, therefore,
can be a potential tool for the derivation of $\log g$.

The atmospheres of VMP stars are characterized by low electron
number densities and low opacities, which are particularly low in
the ultraviolet spectral range. Hence, local thermodynamical
equilibrium (LTE) is not fulfilled in the atmospheric layers where
the lines are formed. As was proven in the past decade, the
surface gravity based on the ionization balance
\ion{Fe}{i}/\ion{Fe}{ii}\, (Thevenin \& Idiart \cite{fe99}, Korn
et al. \cite{Korn03}) and element abundances (for review, see
Asplund \cite{asplund05}) are significantly affected by departures
from LTE. In this study, we select four metal-poor stars observed
with the High Dispersion Spectrograph of the Subaru Telescope and
perform detailed analysis of stellar parameters and element
abundances based on the non-local thermodynamic equilibrium
(non-LTE) line formation for 10 important chemical species,
including \ion{H}{i}, \ion{Ca}{i}, \ion{Ca}{ii}, etc. One of the
stars, HD\,122563, is a giant, and the remaining three, HD\,84937,
BD\,$+3^\circ$740, and BD\,$-13^\circ$3442, are at the hot end of
the stars that evolve on time scales comparable to the Galaxy
lifetime. Similar intrinsically bright objects are mainly found
among VMP stars due to the selection effect. For example, one of
two HMP stars, the Christlieb's star is a giant with $\log g$ =
2.2 (Christlieb et al. \cite{chri04}) and another one has an
effective temperature $\Teff$ = 6180~K (Aoki et al.
\cite{frebel06}). In the selected stars, it is possible to study
the formation of some of the strongest spectral lines, such as
\ion{Ca}{ii}\, $\lambda8498$, \ion{Mg}{i}\, $\lambda5172$ and
$\lambda5183$, which are among the few spectral lines observed in
HMP stars. At [Fe/H] $< -2$, they are presumably of purely
photospheric origin. Thus, the stars of our small sample can serve
as reference stars in studies of VMP and HMP stars.

The paper is organized as follows. Observations of the selected
stars are described in Sect.\,\ref{obs}. Non-LTE calculations are
found in Sect. \,\ref{NLTE}. The solar element abundances necessary
for further differential abundance analysis are derived in
Sect.\,\ref{line_list} from the selected spectral lines. In
Sect.\,\ref{st_param}, we determine effective temperatures from
the wings of Balmer lines and check surface gravities from the
ionization balance \ion{Ca}{i}/\ion{Ca}{ii}. Non-LTE abundances of
8 chemical elements are derived in Sect.\,\ref{Abundance}. For two
stars, HD\,84937 and HD\,122563, both the \ion{Ba}{ii}\, resonance
and subordinate lines are available, and we determine in
Sect.\,\ref{isotope} the fraction of the odd isotopes of barium that gives
an independent estimate of relative contribution of the
$r-$process and main $s-$process to heavy element abundances in
the star. Our recommendations and conclusions are given in
Sect.\,\ref{conclusion}.

\section{Observations and data reductions}\label{obs}

The spectroscopic observations for our program stars were carried
out on Feb. 2, 2002 with the High Dispersion Spectrograph (HDS;
Noguchi et al. \cite{Noguchi}) at the Nasmyth focus of the Subaru
8.2m telescope. Signal-to-noise (S/N) ratio measured per pixel
(0.9km/s) is 200 to 400 for the brightest star HD\,122563 and is
lower, down to 100, for the remaining three stars. We use also
high-quality observed spectra covering \ion{Ca}{ii}\, 3933,
\ion{Ca}{ii}\, 8498, the lines of \ion{Al}{i} and \ion{K}{i} from
the ESO UVESPOP survey (Bagnulo et al. \cite{POP03}) for HD\,84937
and HD\,122583 and from the ESO/UVES archive (Program ID:
67.D-0106 by P. Nissen; 67.D-0439 and 68.D-0094 by F. Primas) for
BD\,$+3^\circ$740 and BD\,$-13^\circ$3442. The Balmer lines
H$_\alpha$ and H$_\beta$ in HD\,84937 and BD\,$+3^\circ$740 are
studied using observational data obtained by Klaus Fuhrmann with
the fiber optics Cassegrain echelle spectrograph FOCES at the 2.2m
telescope of the Calar Alto Observatory in 1999 and 1995. The high
quality of the FOCES data reduction accuracy has been emphasized
by Korn (\cite{Korn02}) who presents evidence for his claim that
systematic errors of the effective temperatures derived from the
Balmer line wing fits due to data reduction (rectification and
flatfielding) are below 50~K. This holds for HD\,84937. For the
BD\,$+3^\circ$740 spectrum, the error may reach a slightly higher
value ($\sim$70~K), because before 1997 all spectra were degraded
by mode transfer noise in the optical fibers (Grupp \cite{grupp})
that did not allow an S/N $> 150 ... 200$. Characteristics of
observed spectra are summarized in Table\,\ref{obs_data}.

The Subaru and ESO/UVES spectra were reduced with a standard
MIDAS/ECHELLE package for order identification, background
subtraction, flat-fielding, order extraction, and wavelength
calibration. Bias, dark current, and scatted light correction are
included in the background subtraction. The spectrum was then
normalized by a continuum function determined by fitting a spline
curve to a set of pre-selected continuum windows estimated from
 the Kitt Peak Solar Atlas (Kurucz et al. \cite{Atlas}).

\begin{table*}[!t]
\setlength{\tabcolsep}{3.5mm}
\begin{center}
\caption{Characteristics of observed spectra.} \label{obs_data}
\begin{tabular}{llccl}
\hline
Telescope /    & \multicolumn{1}{c}{Spectral range (\AA)}   & R & S/N & Objects \\
spectrograph  &   &   &     &  \\
\hline
Subaru / HDS    & 4120 - 5430, 5520 - 6860 & 90000 & $\ge 200$ & HD\,122563 \\
              &    &       & $\ge 100$ & HD\,84937, BD\,$+3^\circ$740, BD\,$-13^\circ$3442 \\
VLT2 / UVES     & 3850 - 8550            & 80000 & $\ge 200$ & HD\,122563, HD\,84937 \\
VLT2 / UVES     & 3730 - 4990, 4760 - 6840, & 48000, & $\ge 200$ & BD\,$+3^\circ$740, BD\,$-13^\circ$3442 \\
              & 6600 - 10600           &  60000      &           & \\
2.2-m / FOCES   & 4500 - 6700            & 60000 & $\ge 100$ & HD\,84937 \\
              &                      & 35000 & $\ge 100$ & BD\,$+3^\circ$740 \\
\hline
\end{tabular}
\end{center}
\end{table*}

\section{Non-LTE calculations} \label{NLTE}

All the investigated elements are assumed to be trace elements,
which means that we obtain statistical equilibrium (SE)
populations for each of them while keeping the atmospheric
structure fixed. In the stellar parameter range we are concerned
with, such approach is justified both for metals and for hydrogen.
Hydrogen is not an important continuous opacity source compared to
the H$^-$ ions at temperature below 7000~K. Although it becomes a
dominant electron donor in the atmospheres of metal-poor stars
([Fe/H] $< -1$), our non-LTE calculations show that the ionization
balance between \ion{H}{ii} and \ion{H}{i} is close to
thermodynamic equilibrium (TE) in the layers where the Balmer line
wings form. For instance, the population ratio
\ion{H}{ii}/\ion{H}{i} deviates by no more than 1\%\, from the TE
value below $\log \tau_{5000} = -2.5$ in the model with $\Teff =
6390$K, $\log g = 3.88$, and [Fe/H] = --2.66. Based on the
self-consistent non-LTE modelling of a solar-type star, Short \&
Hauschildt (\cite{short}) conclude that the non-LTE effects of all
the light metals (including Na, Mg, Al, K, and Ca among others) on
the model structure and flux distribution are small.

Calculations are performed with the plane-parallel, homogeneous,
LTE, and blanketed model atmospheres computed for individual
stellar parameters using the code MAFAGS (Fuhrmann et al.
\cite{Fuhr1}). For the program metal-poor stars,
$\alpha$-enhancement is accounted for with the abundances of Mg
and Ca determined in this study, the Si abundance that follows the
Mg one, and the oxygen abundance taken from the literature. We
assume [O/Fe] = 0.5 where no other data is available. It is worth
noting that oxygen in cool stellar atmospheres plays a minor role
as a donator of free electrons and as opacity source, and the
uncertainty of its abundance does not affect the calculated
atmospheric structure.

In order to solve the coupled radiative transfer and statistical
equilibrium equations for metals, we use a revised version of the
DETAIL program (Butler \& Giddings \cite{detail}) based on the
accelerated lambda iteration, which follows the efficient method
described by Rybicki \& Hummer (\cite{rh91}, \cite{rh92}). Non-LTE
computations for \ion{H}{i}\, are based on the complete
linearization method as described by Auer \& Heasley (\cite{AH}).
We apply the code NONLTE3 originally treated by Sakhibullin
(\cite{Sa}) and advanced later (Kamp et al. \cite {nonlte3}). The
departure coefficients are then used to compute the synthetic line
profiles. The metal line list has been extracted from Kurucz'
(\cite{cdrom18}) compilation.

\subsection{Non-LTE line formation for \ion{H}{i}}

\subsubsection{Model atom of \ion{H}{i}}

We consider non-LTE line formation for \ion{H}{i}\, in order to
provide diagnostics for the Balmer lines H$_\alpha$ and H$_\beta$ that are
used for an accurate temperature determination in cool stars. The
model atom of \ion{H}{i}\, includes levels with principal quantum
numbers up to n $\le$ 19 and energies adopted from Wiese et al.
(\cite{WSG}). Transition probabilities are taken from the Vienna
Atomic Line Data base (Kupka et al. \cite{vald}) and if not
available they are computed using the approximate formula of Bethe
\& Salpeter (\cite{bethe}) as implemented by Johnson
(\cite{johnson}). Photoionization cross-sections are evaluated
using the exact expression for the hydrogen atom. Collisional rates include contributions from inelastic collisions with electrons and hydrogen atoms. For electron-impact excitation, the
$R-$matrix calculations of Przybilla \& Butler
(\cite{nlte1}) are used for the transitions between the energy levels with
$n \le 7$ and the approximation formula of Johnson
(\cite{johnson}) for the remainder. For electron-impact ionization, we
 apply the Seaton formula as described by Mihalas (\cite{mihalas}).
The data on inelastic collisions with neutral H particles available in literature for \ion{H}{i} have been recently investigated by Barklem (\cite{barklem_h}). He concludes that
they all are of questionable quality. Here we use the Drawin's
(\cite{D68}) formula as implemented by Steenbock \& Holweger (\cite{hyd}).
Since it provides only an order of magnitude estimate, we constrain the efficiency of
hydrogenic collisions empirically. The rates calculated using this formula were multiplied by
a scaling factor \kH\ = 0 (no hydrogenic collisions), 0.1, 1, and 2 in order to
make the effective temperatures derived from the two Balmer lines H$_\alpha$ and H$_\beta$ in the selected stars consistent. Our best estimate \kH\ = 2 agrees with that of Przybilla \& Butler
(\cite{nlte2}).

Radiation transfer is treated explicitly for all Balmer and
Paschen lines, for the first seven Lyman lines, for every
bound-bound transition with $n_{low} =$ 4 -- 8 and $n_{up} \le
13$, and for the four continua including Lyman, Balmer, Paschen
and Brackett ones, in total, for 70 transitions. Radiative rates
for the remaining transitions are evaluated using LTE mean
intensities obtained from the formal solution of the radiation
transfer equation. For the transitions arising between the energy
levels with $n \le 8$, the line absorption profile is computed in
detail and includes Stark broadening, radiative damping,
self-broadening, Doppler broadening (both thermal and turbulent),
and fine structure as implemented by Barklem \& Piskunov
(\cite{hlinop}) in their subroutine HLINOP.
 For the remaining transitions, Doppler profiles are assumed.

\subsubsection{Departures from LTE for \ion{H}{i}}

\begin{figure}
\resizebox{88mm}{!}{\includegraphics{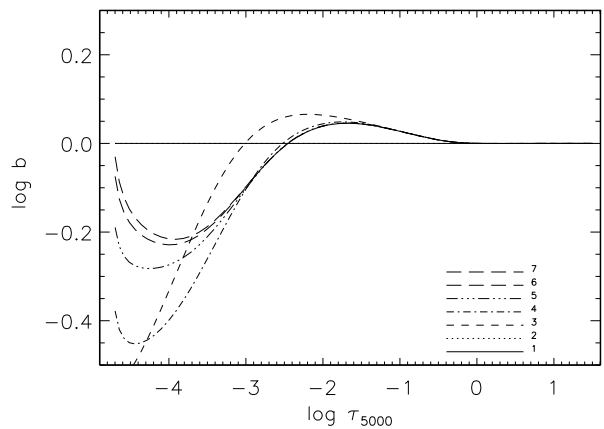}}
\caption[]{Departure coefficients $\log b_i$ for the first seven
levels of \ion{H}{i}\, in the model atmosphere of
BD$+3^\circ$740.} \label{h_bf}
\end{figure}

Our calculations show that the behavior of the departure
coefficients $b_i = n_i^{\rm NLTE}/n_i^{\rm LTE}$ and the line
source functions are similar for all the models representing
the atmospheres of the Sun and selected stars. Here, $n_i^{\rm NLTE}$
and $n_i^{\rm LTE}$ are the statistical equilibrium and thermal
(Saha-Boltzmann) number densities, respectively. In Fig.\,\ref{h_bf}, the departure coefficients of the lowest 7 levels of
\ion{H}{i}\, are shown as a function of continuum optical depth
$\tau_{5000}$ referring to $\lambda = 5000$\AA\ in the model
atmosphere of BD$+3^\circ$740. The ground state and the first
excited level ($n = 2$) keep their TE level
populations throughout the atmosphere.
Departures from LTE for the $n = 3$ level are controlled by
H$_\alpha$. In the layers, where the continuum optical depth drops
below unity, H$_\alpha$ serves as the pumping transition resulting
in a slight overpopulation (up to $\simeq5\%$) of the upper level.
But starting from $\log \tau_{5000} \simeq -3$ and further out,
photon escape from H$_\alpha$ wings causes an underpopulation of
the $n = 3$ level. The levels with $n \ge 4$ stay in
nearly detailed balance relative to each other.

For the Balmer lines of particular interest, H$_\alpha$ and
H$_\beta$, non-LTE leads to slight weakening of the
core-to-wing transition compared to the LTE case and significant
strengthening of the core. The core-to-wing
transition is of particular importance for temperature
determinations, because this part of profile is
sensitive to $\Teff$ variations.

We find that classical theoretical model atmospheres fail to
reproduce a half-width of the H$_\alpha$ and H$_\beta$ lines in
the Sun (Fig.\,\ref{ha_sun} for H$_\alpha$) and cool stars
(Fig.\,\ref{h_122563} and \ref{h_g84_29}). Similar conclusion is
drawn by Przybilla \& Butler (\cite{nlte2}) and Fuhrmeister et al.
(\cite{Fuhrmeister}) for H$_\alpha$ in the Sun. Therefore the line
core within 0.7\AA\, to 1.5\AA, with the exact number depending on
stellar parameters, is not included in the fit. In the entire
range of stellar parameters we are concerned with, we obtain weak
non-LTE effects for the H$_\beta$ profile beyond the core
independent of the applied scaling factor \kH. For example, for
the model representing the atmosphere of BD\,$+3^\circ$740
($\Teff$ = 6340~K, $\log g$ = 3.90, [Fe/H] = -2.65), a difference
in the LTE and non-LTE (\kH\, = 0.1) profiles equals 0.1\%\, of
the continuum flux at $\Delta\lambda$ = 1.7\AA\, from the line
center and rapidly decreases with increasing $\Delta\lambda$. In
fact, non-LTE effects can be neglected deriving the effective
temperature from H$_\beta$. For H$_\alpha$, the corresponding
difference equals 2.7\%\, of the continuum flux at $\Delta\lambda$
= 1.7\AA\, and becomes smaller than 0.1\%\, only beyond
$\Delta\lambda$ = 6.7\AA. A difference in the non-LTE H$_\alpha$
profiles computed with \kH\, = 1 and 2 is smaller than 0.1\%\, of
the continuum flux everywhere beyond $\Delta\lambda$ = 1.7\AA.
Both profiles are 0.7\%\, stronger compared to that computed with
\kH\, = 0.1 at $\Delta\lambda$ = 1.7\AA. The three profiles agree
within 0.1\%\, beyond $\Delta\lambda$ = 3.2\AA. In
Sect.\,\ref{bd3_740}, we inspect an influence of the \kH\,
variation on the derived $\Teff$ of BD\,$+3^\circ$740. Similar
effects are expected for other program stars.

\subsubsection{H$_\alpha$ in the solar spectrum}

Hereafter, the theoretical profiles of H$_\alpha$ and H$_\beta$
are computed as accurately as possible, including a convolution of
the profiles resulting from the different broadening mechanisms.
The Stark broadening profiles from Vidal et al. (\cite{vcs70,
vcs73}) are employed. For self-broadening, we use two recipes. The
first one is based on the Ali \& Griem (\cite{AG66}) theory
(hereafter $AG$) and the second one on the self-broadening
formalism of Barklem et al. (\cite{BPOa}, hereafter $BPO$). We
apply the revised subroutine HYDLINE incorporated in the SIU code.

\begin{figure}
\resizebox{88mm}{!}{\includegraphics{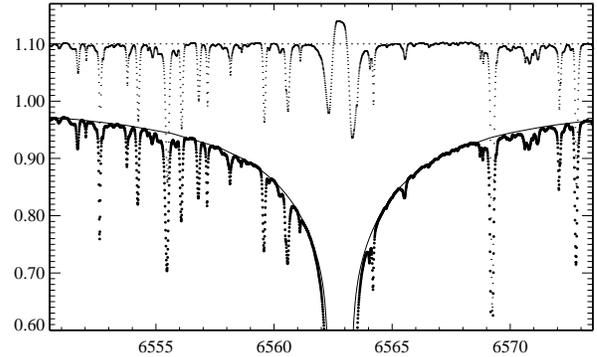}}
\caption[]{ Theoretical non-LTE flux profile (continuous line) of
the Balmer H$_\alpha$ line computed for $\Teff$ = 5780~K using the
$AG$ self-broadening recipe compared to the observed spectrum of
the Kurucz et al. (1984) solar flux atlas (bold dots). The
differences between observed and calculated spectra (O - C) are
presented in the upper part of figure. } \label{ha_sun}
\end{figure}

Non-LTE modelling of H$_\alpha$ has been tested by comparing with
solar flux observations taken from the Kitt Peak Solar Atlas
(Kurucz et al. \cite{Atlas}). We use the grid of the MAFAGS model
atmospheres with various $\Teff$ but with the fixed values of
$\log g = 4.44$ and solar metal abundance. The line core within
$\pm1.1\AA$ is not included in the fit. We obtain smaller non-LTE
effects for H$_\alpha$ in the Sun compared to that found in our
program metal-poor stars. The best non-LTE fit is achieved for
$\Teff(AG)$ = 5780~K when the $AG$ recipe is applied
(Fig.\,\ref{ha_sun}) and for a lower temperature $\Teff(BPO)$ =
5720~K in the case of the $BPO$ recipe. In each case, the LTE
calculated wings for the same temperature are slightly deeper (by
up to 0.3\%) than the observed profile, indicating a 30~K lower
effective temperature value. The H$_\beta$ wings are strongly
blended in the solar spectrum. They are satisfactorily fitted
using $\Teff(AG)$ = 5780~K.

\subsection{Non-LTE calculations for metals}

For metals, our present investigation is based on the non-LTE
methods treated in our earlier studies and documented in a number
of papers, where atomic data and the problems of line formation
have been considered in detail. Table~\ref{papers} lists the
investigated atoms and cites the related papers. For \ion{Ca}{ii},
the advanced effective collision strengths are used from the
recent $R-$matrix calculations of Mel\'endez et al.
(\cite{ca2_bautista}).

\begin{table}[!th]
\caption{Atomic models used in this study where NL is the number
of levels in the included species$^1$.}\label{papers}
\begin{center}
\begin{tabular}{llll}
\hline
Atom &  & NL  & Reference \\
\hline
 Sodium & \ion{Na}{i} & 58 & Baum\"uller et al.
(\cite{nlte_na}), \\
       &              &     & Gehren et al. (\cite{mg_c6}),\\
       &              &     & Shi et al. (\cite{shi04})\\
Magnesium &\ion{Mg}{i} &85 & Zhao et al. (\cite{nlte_mg}),\\
       &              &     & Gehren et al. (\cite{mg_c6})\\
Aluminium &\ion{Al}{i} &59 & Baum\"uller et al.
(\cite{nlte_al}) \\
Potassium & \ion{K}{i} & 67 & Zhang et al. (\cite{nlte_ka,
nlte_kb}) \\
Calcium & \ion{Ca}{i} & 63 & Mashonkina et al.
(\cite{mash_ca}) \\
        & \ion{Ca}{ii} & 37 & \\
Strontium & \ion{Sr}{ii} & 40 & Belyakova \& Mashonkina
(\cite{sr}), \\
       &              &        & Mashonkina \& Gehren
(\cite{eubasr}) \\
Barium & \ion{Ba}{ii} & 35 &  Mashonkina et al. (\cite{Mash99}) \\
Europium & \ion{Eu}{ii} & 63$^2$ & Mashonkina (\cite{MLeu}), \\
         &               &    & Mashonkina \& Gehren
(\cite{euba}), \\
         &               & & Mashonkina \& Vinogradova
         (\cite{eu_new}) \\
\hline
\end{tabular}
\end{center}
 \ $^1$ Each model atom is closed
with the ground state of the next ionization stage, \\
 \ $^2$ combined levels based on 176 energy levels of
\ion{Eu}{ii}\, known from laboratory measurements.
\end{table}

We note that the most accurate atomic data are used in
our non-LTE calculations. For minority species such as
\ion{Mg}{i}, \ion{Al}{i}, and \ion{Ca}{i}, where departures from
LTE are mainly caused by superthermal radiation of non-local
origin below the thresholds of low excitation levels,
photoionization cross-sections are from the Opacity Project
calculations (Seaton et al. \cite{OP}). Their accuracy is
estimated at the level of 10\%. Such small uncertainty translates
to the abundance error of no more than 0.01~dex. For the
collision-dominated atoms \ion{Na}{i}\, and \ion{K}{i}\, and the
majority species such as \ion{Ca}{ii}, \ion{Sr}{ii}, \ion{Ba}{ii},
and \ion{Eu}{ii}, the main source of uncertainties is collisional
data. In SE calculations, we account for inelastic collisions both
with electrons and neutral H particles. Hydrogen collisions are computed using the formula
of Steenbock \& Holweger (\cite{hyd}) with a scaling factor
\kH\, found empirically from their different influence on the
different lines of a given atom in solar and stellar spectra. We
apply \kH\, = 0 (no hydrogenic collisions) for \ion{Sr}{ii},
\ion{Ba}{ii}, and \ion{Eu}{ii}\, as estimated by Mashonkina \& Gehren
(\cite{euba, eubasr}) and Mashonkina \& Vinogradova (\cite{eu_new}), \kH\, = 0.002 for \ion{Al}{i}\, and
\kH\, = 0.05 for \ion{Na}{i}\, according to Gehren et al.
(\cite{mg_c6}), \kH\, = 0.05 for \ion{K}{i}\, (Zhang et al.
\cite{nlte_ka, nlte_kb}), and \kH\, = 0.1 for \ion{Mg}{i}\, and
\ion{Ca}{i/ii}\, as recommended by Mashonkina et al.
(\cite{mash_ca}).

\section{Line list and solar abundances for a differential
analysis}\label{line_list}

The spectral lines used in our study are listed in
Table\,\ref{lines}. Hyperfine structure (HFS) and/or isotope
structure (IS) is taken into account when necessary with the data
from Moore (\cite{moore72}, IS for \ion{Mg}{i} $\lambda4703$,
$\lambda5528$, and $\lambda5711$), N\"ortersh\"auser et al.
(\cite{ca_is}, IS for \ion{Ca}{ii}\, $\lambda8498$), McWilliam et
al. (\cite{mcw95}, \ion{Sr}{ii}\, $\lambda4215$), McWilliam
(\cite{mcw98}, \ion{Ba}{ii}\, $\lambda4554$), and Lawler et al.
(\cite{eu_lawler}, \ion{Eu}{ii}\, $\lambda4129$). For Mg, Ca, Sr,
and Eu, we use the fractional isotope abundances corresponding to
the solar system matter (Anders \& Grevesse \cite{AG}). For
barium, the even-to-odd isotope abundance ratio was allowed to
vary, and its value is determined in Sect.\,\ref{isotope} for the
two stars.

\begin{table}[!th]
\caption{Atomic data and solar abundances for the investigated
lines.}\label{lines}
\begin{center}
\begin{tabular}{lcccrcrc}
\hline \noalign{\smallskip}
 Ion & $\lambda,\AA$ & E$_{exc}$ & $\log gf$ & Ref &
 $\log C_6$ & Ref & $\eps{\odot}$  \\
\noalign{\smallskip} \hline \noalign{\smallskip}
\ion{Na}{i} & 5889.96 & 0.00 & ~0.11 & 2 & --31.60 & 14 & 6.28  \\
 & 5895.93 & 0.00 & --0.19 & 2 & --31.60 & 14 & 6.28  \\
\ion{Mg}{i} & 4571.10$^1$ & 0.00 & --5.62 & 2 & --31.96 & 14 & 7.58 \\
 & 4702.99$^1$ & 4.33 & --0.44 & 2 & --29.71 & 14 & 7.49 \\
 & 5528.41$^1$ & 4.33 & --0.50 & 2 & --30.20 & 14 & 7.53 \\
 & 5711.09$^1$ & 4.33 & --1.72 & 2 & --29.89 & 14 & 7.55 \\
 & 5172.68 & 2.70 & --0.45 & 3 & --30.88 & 14 & 7.56   \\
 & 5183.60 & 2.70 & --0.24 & 3 & --30.88 & 14 & 7.57  \\
\ion{Al}{i} & 3961.53 & 0.01 & --0.34 & 2 & --31.20 & 14 & 6.44 \\
\ion{K}{i} & 7664.92 & 0.00 & ~0.13 & 4 & --31.00 & 5 &5.12\\
 & 7698.98 & 0.00 & --0.17 & 4 & --31.00 & 5 & 5.12  \\
\ion{Ca}{i} & 4425.44  & 1.87& $-$0.36& 6 &  $-$30.90 & * & 6.41  \\
 & 4578.56  & 2.51& $-$0.70& 7 &  $-$30.30 & 16 & 6.37  \\
 & 5261.71  & 2.51& $-$0.58& 7 &  $-$30.86 & 16 & 6.44  \\
 & 5349.47  & 2.70& $-$0.31& 7 &  $-$31.45 & 16 & 6.41  \\
 & 5588.76  & 2.51& ~0.36  & 7 &  $-$31.39 & 16 & 6.28 \\
 & 5590.12  & 2.51& $-$0.57& 7 &  $-$31.39 & 16 & 6.39 \\
 & 5857.45  & 2.92& ~0.24  & 7 &  $-$30.61 & 16 & 6.36 \\
 & 6122.22  & 1.88& $-$0.32& 6 &  $-$30.30 & 15 & 6.26 \\
 & 6162.17  & 1.89& $-$0.09& 6 &  $-$30.30 & 15 & 6.28 \\
 & 6166.44  & 2.51& $-$1.14& 7 &  $-$30.48 & 16 & 6.42 \\
 & 6439.07  & 2.51& ~0.39  & 7 &  $-$31.58 & 16 & 6.27 \\
 & 6471.66  & 2.51& $-$0.69& 7 &  $-$31.58 & 16 & 6.34 \\
 & 6493.78  & 2.51& $-$0.11& 7 &  $-$31.58 & 16 & 6.31 \\
 & 6499.65  & 2.51& $-$0.82& 7 &  $-$31.58 & 16 & 6.38 \\
 & 6449.81  & 2.51& $-$0.50& 7 &  $-$31.45 & 16 & 6.34 \\
\ion{Ca}{ii}& 3933.66 & 0.00& ~0.10& 8 & $-$31.72 & 17 & 6.35  \\
  & 8498.02$^1$ & 1.69& $-$1.42& 8 & $-$31.51 & 17 & 6.35 \\
\ion{Fe}{ii} & 4508.29 & 2.84& --2.32& 9 & $-$32.00 & 18& 7.44  \\
 & 4582.84 & 2.83& --3.14& 9 & $-$32.03 & 18& 7.43  \\
 & 4923.93 & 2.88& --1.36& 9 & --32.05  & 18& 7.41  \\
 & 5018.44 & 2.88& --1.23& 9 & --32.05  & 18& 7.45  \\
 & 5197.57 & 3.22& --2.24& 9 & --32.05  & 18& 7.46  \\
 & 5234.62 & 3.21& --2.14& 9 & --32.05  & 18& 7.41  \\
 & 5264.81 & 3.22& --3.13& 13& --32.03  & 18& 7.53  \\
 & 5325.56 & 3.21& --3.26& 12& --32.05  & 18& 7.56  \\
 & 5425.25 & 3.19& --3.30& 12& --32.05  & 18& 7.51  \\
 & 6247.56 & 3.87& --2.43& 13& --32.00  & 18& 7.57  \\
 & 6456.38 & 3.89& --2.18& 13& --32.00  & 18& 7.59  \\
\ion{Sr}{ii} & 4215.53$^1$ & 0.00& $-$0.17& 10& $-$31.80 & * & 2.92 \\
\ion{Ba}{ii} & 4554.03$^1$ & 0.00& ~0.16  & 10& $-$31.65 & * & 2.21 \\
             & 6496.90     & 0.60 & --0.38& 10& $-$31.28 & 17& 2.21 \\
\ion{Eu}{ii} & 4129.72$^1$ & 0.00& ~0.22  & 11& --32.08  & 19 & 0.56 \\
\noalign{\smallskip} \hline
\end{tabular}
\end{center}
Excitation energy E$_{exc}$ is given in eV; \\
 \ $^1$ IS and HFS components are taken into account; \\
Ref.: 2 - NIST database; 3 - Aldenius et al. (\cite{mg_ald}); 4 -
Butler (\cite{butler2000}); 5 - Zhang et al. (\cite{nlte_ka});
6 - Smith \& O'Neil (\cite{ca_75}); 7 - Smith \&
Raggett (\cite{ca_fij}); 8 - Theodosiou (\cite{ca3933}); 9 -
Landstreet (\cite{fe2_land}); 10 - Reader et al. (\cite{WM}); 11 -
Lawler et al. (\cite{eu_lawler}); 12 - Moity (\cite{moity});
13 - Raassen \& Uylings (\cite{fe2_raas}); 14 - Gehren et al.
(\cite{mg_c6}); 15 - Anstee \& O'Mara (\cite{omara_sp}); 16 -
Smith (\cite{ca_81}); 17 - Barklem \& O'Mara (\cite{omara_ion});
18 - Barklem \& Aspelund-Johansson (\cite{baj05}); 19 - Kurucz
(\cite{cdrom18}); * - solar line profile fitting. \\
\end{table}

Based on non-LTE line formation, we determine the solar element
abundance from each individual line. The exception is the
\ion{Fe}{ii}\, lines, which are analyzed assuming LTE. The solar
flux profiles from the Kurucz et al. (\cite{Atlas}) Atlas are
fitted using the fixed oscillator strengths and van der Waals
damping constants collected in Table\,\ref{lines}. The profile of
the strongest Ca line \ion{Ca}{ii}\,$\lambda3933$ cannot be fitted
although the best available atomic data are used. We assume that
its core and inner wings are influenced by the chromospheric
temperature rise and non-thermal depth-dependent chromospheric
velocity field that is not part of the MAFAGS model. For this
line, we adopt the mean Ca abundance derived from the
\ion{Ca}{i}\, lines.

We find that the mean solar abundances derived in this study agree
within the error bars with the corresponding meteoritic abundances
of Asplund et al. (\cite{met05}). This gives confidence in the
reliability of our results obtained in analysis of stellar
spectra. Although analysis is made line-by-line differentially
with respect to the Sun, it is not strictly differential. A line
which is measurable in the selected metal-poor stars, is strong in
the solar spectrum, and the derived solar element abundance
depends not only on its $gf-$value but also on van der Waals
damping parameter.

\section{Stellar parameters}\label{st_param}

Our results both for stellar parameters and element abundances are
based on line profile analysis. In order to compare with
observations, computed synthetic profiles are convolved with a
profile that combines instrumental broadening with a Gaussian
profile and broadening by macroturbulence with a radial-tangential
profile. For a given star, the macroturbulence value $\Vmac$ was
allowed to vary by $\pm$0.5\,\kms\ (1$\sigma$). Our strategy in
this section is as follows.
\begin{enumerate}
\item Stellar effective temperature is determined from the
hydrogen H$_\alpha$ and H$_\beta$ line wing fitting based on
non-LTE line formation. As was shown by Fuhrmann et al.
(\cite{Fuhr93}), the Balmer line profiles are strongly temperature
sensitive, while the variation with gravity and the metal
abundance is rather small. H$_\alpha$ in particular is very
insensitive to $g$ and [M/H]. This is the basic justification for
using the Balmer lines as a temperature indicator. \item Next, the
surface gravity is derived from the {\sc Hipparcos} parallax with
a mass determined from the tracks of VandenBerg et al.
(\cite{isohrone}). The exception is BD\,$-13^\circ$3442 with no
{\sc Hipparcos} parallax available. \item The ionization balance
between \ion{Ca}{i}\, and \ion{Ca}{ii}\, is checked for the fixed
temperature and current estimate of $\log g$. We denote the mean
abundance from the subordinate lines of \ion{Ca}{i}\, as the CaI
abundance and from \ion{Ca}{ii} $\lambda3933$ and \ion{Ca}{ii}
$\lambda8498$ as the CaII abundance. If (CaI - CaII) exceeds the
abundance error, the gravity is revised. We note that the two
\ion{Ca}{ii} lines behave oppositely with varying gravity. The
resonance line has the van der Waals broadened wings even at
[Ca/H] $< -2$, and the derived Ca abundance decreases/increases
with increasing/decreasing gravity. The IR line is strengthened
with decreasing $\log g$ due to decreasing the H$^-$ continuous
absorption and due to amplified departures from LTE for
\ion{Ca}{ii}. This gives an additional opportunity to constrain
stellar gravity from a comparison of the element abundances from
the two \ion{Ca}{ii} lines. \item Simultaneously, the iron
abundance is obtained from the \ion{Fe}{ii}\, lines assuming LTE,
and the abundances of $\alpha-$process elements Mg and Ca are
determined from non-LTE analysis of the \ion{Mg}{i}\, and
\ion{Ca}{i}\, lines. \item A microturbulence velocity $\Vmic$ is
derived from the strongest \ion{Fe}{ii}, \ion{Ca}{i}, and
\ion{Mg}{i}\, lines sensitive at given stellar parameters to a
variation in $\Vmic$. \end{enumerate}

If any of the obtained parameters, either $\Teff$, or $\log g$, or
[Fe/H], or $\alpha-$enhancement, deviates from the one adopted
originally, the model atmosphere is recalculated and the steps 1
through 5 are repeated. Thus, stellar parameters are determined
iteratively. We describe below how it is made for the individual
stars and order them by the level of the accuracy of stellar
parameters $\Teff$ and $\log g$ available in the literature, from
the higher accuracy to the lower one. The final stellar parameters
and element abundances are presented in Table\,\ref{startab}. The
mean element abundance is given where more than one line is used
in analysis, and the error bars quoted in Table\,\ref{startab} is
computed as the standard deviation $\sigma =
\sqrt{\Sigma(\overline{x}-x_i)^2/(n-1)}$.

\subsection{\rm HD\,84937}

This star with a bright magnitude of 8.28 is among the best
studied halo stars. Its effective temperature was determined many
times (for review, see Korn et al. \cite{Korn03}) both spectroscopically from Balmer line wing fitting and
using the infrared flux method (IRFM) with surprisingly consistent
results. According to the LTE analysis of Korn et al.
(\cite{Korn03}), $\Teff$ (Balmer) = 6346~K, while Alonso et al.
(\cite{alonso96}) and Mel\'endez \& Ram\'irez (\cite{melendez04})
give $\Teff$(IRFM) = 6330~K and 6345~K, correspondingly. The {\sc
Hipparcos} parallax, $\pi = 12.44\pm1.06$~mas is known with an
accuracy better than 10\%.

The Balmer line profiles are computed applying the two
self-broadening recipes, $AG$ and $BPO$. In non-LTE, we obtain
$\Teff(AG)$ = 6380~K from H$_\alpha$ and 6350~K from H$_\beta$.
Based on a $S/N$ ratio of the observed spectrum and a sensitivity
of the Balmer lines to a variation of $\Teff$, we estimate the
uncertainty of $\Teff$ arising from the profile fitting as 30~K
for H$_\alpha$ and 60~K for H$_\beta$. The core within
$\pm$1~\AA\, for H$_\alpha$ and $\pm$0.7~\AA\, for H$_\beta$ is
not included in the fit. A mean temperature, $\Teff$ = 6365~K, is
adopted as a final value. In the case of the self-broadening from
Barklem et al. (\cite{BPOa}), a mean temperature $\Teff(BPO)$ =
6300~K is derived from two Balmer lines. The maximum error of
$\Teff$ then is, probably, not larger than 70\,K.

A trigonometric gravity was carefully calculated by Korn et al.
(\cite{Korn03}), $\log g = 4.00 \pm 0.12$. For a temperature
$\Teff$ = 6365~K obtained in this study, $\log g$ increases by
less than 0.01. For the model with $\Teff$/$\log g$ = 6365/4.00,
we find a good agreement between the non-LTE CaI abundance based
on 14 lines of \ion{Ca}{i}\, and the non-LTE CaII abundance from
two \ion{Ca}{ii}\, lines: (CaI - CaII) = 0.01. In contrary, an
agreement between different lines is destroyed at the LTE
assumption (see Table\,\ref{startab}).

The error of the abundance difference (CaI - CaII) is mainly
determined by the error of the CaII abundance, because the error
of the CaI abundance is very small for the program stars. It
constitutes $\sigma/\sqrt{n}$ where $n = 15$ to 10 for different
stars, and $\sigma$ does not exceed 0.06~dex
(Table\,\ref{startab}). For \ion{Ca}{ii} $\lambda3933$ in
different stars, the abundance error is mainly due to the
uncertainties of the van der Waals damping constant and the
profile fitting, while the uncertainty of microturbulence velocity
dominates the abundance error for \ion{Ca}{ii} $\lambda8498$. The
van der Waals damping constants predicted by the perturbation
theory of Barklem \& O'Mara (\cite{omara_ion}) have the
uncertainty $\Delta\log C_6$ = 0.05 -- 0.18~dex according to
Barklem \& Aspelund-Johansson (\cite{baj05}) and Barklem
(\cite{barklem06}). For \ion{Ca}{ii} $\lambda3933$ in the program
stars, this translates to the abundance error of up to 0.06~dex.

For \ion{Ca}{ii} $\lambda3933$ in HD\,84937, only the line wings
are fitted because the observed Doppler core is very broad and
most probably affected by the chromospheric temperature rise and
non-thermal and depth-dependent chromospheric velocity fields
similar to solar ones. The abundance error is estimated as
0.06~dex. For \ion{Ca}{ii} $\lambda8498$, $\Delta\Vmic = 0.2$~\kms
\ leads to $\sigma =$0.04~dex. Thus, the differences of LTE
abundances (CaII ($\lambda8498$) - CaI) and (CaII ($\lambda8498$)
- CaII ($\lambda3933$)) exceed their three to two errors.

The iron abundance [Fe/H] = --2.15 and microturbulence velocity $\Vmic$ = 1.6~\kms
are determined from 10 \ion{Fe}{ii}\, lines requiring that the
derived abundances do not depend on line strength. The
$\alpha-$enhancement of the model is defined by the non-LTE Mg
abundance, [Mg/Fe] = 0.39.

\begin{table}[!th]
\caption{Stellar parameters and element abundances of the selected
stars where $\Vmic$ is given in \kms} \label{startab}
\begin{center}
\begin{tabular}{lcccc}
\hline
HD/BD           & 122563     & 84937     & $+3^\circ$740    & $-13^\circ$3442 \\
\hline
 \ $\Teff$,K    & 4600       & 6365      & 6340             &  6390 \\
 \ $\log g$     & 1.50\scriptsize{$\pm$0.2}& 4.00\scriptsize{$\pm$0.12} & 3.90\scriptsize{$\pm$0.15} &  3.88\scriptsize{$\pm$0.15} \\
 \ [Fe/H]    & --2.53\scriptsize{$\pm$0.02}& --2.15\scriptsize{$\pm$0.04}&--2.65\scriptsize{$\pm$0.01}& --2.66\scriptsize{$\pm$0.01} \\
 \ $\Vmic$      & 1.9        & 1.6       & 1.4              & 1.4 \\
\hline
 & \multicolumn{4}{l}{NLTE} \\
 \ [Mg/H]      &--2.22\scriptsize{$\pm$0.06}& --1.76\scriptsize{$\pm$0.04}& --2.35\scriptsize{$\pm$0.03}& --2.26\scriptsize{$\pm$0.05}  \\
 \ [Na/H]      & --2.93\scriptsize{$\pm$0.01} & --2.48\scriptsize{$\pm$0.02} & --3.22\scriptsize{$\pm$0.03}   & --2.82\scriptsize{$\pm$0.01} \\
 \ [Al/H]      &--3.09 & --2.21 & & --2.61 \\
 \ [CaI/H] & --2.32\scriptsize{$\pm$0.06}& --1.72\scriptsize{$\pm$0.04}& --2.19\scriptsize{$\pm$0.05}& --2.12\scriptsize{$\pm$0.06}  \\
 \ [CaII/H] & & & & \\
 ~~($\lambda$3933)& --2.42 & --1.70 & --2.16 & --2.15 \\
 ~~($\lambda$8498)& --2.36 & --1.76 & --2.14 & --2.06 \\
 \ [K/H]       &--2.60\scriptsize{$\pm$0.04} &  --1.87\scriptsize{$\pm$0.05} & --2.37\scriptsize{$\pm$0.04} & --2.28 \\
 \ [Ba/H] & --3.54 & --2.15    & 2.85 & --2.89 \\
 \ [Sr/H] & --2.93 & --2.18    & --2.62 & --2.35 \\
 \ [Eu/H] & --3.04 & --1.45  & & \\
 & \multicolumn{4}{l}{LTE} \\
 \ [Mg/H]      & --2.39\scriptsize{$\pm$0.10}& --1.88\scriptsize{$\pm$0.03}& --2.40\scriptsize{$\pm$0.05}& --2.29\scriptsize{$\pm$0.03}  \\
 \ [Na/H]      & --2.51\scriptsize{$\pm$0.04} & --2.02\scriptsize{$\pm$0.03} & --3.03\scriptsize{$\pm$0.02} & --2.58\scriptsize{$\pm$0.08} \\
 \ [Al/H]      & --3.13 & --3.04 & & --3.47 \\
 \ [CaI/H] & --2.52\scriptsize{$\pm$0.10}& --1.76\scriptsize{$\pm$0.04}& --2.32\scriptsize{$\pm$0.07}& --2.27\scriptsize{$\pm$0.06}  \\
 \ [CaII/H] & & & & \\
 ~~($\lambda$3933)& --2.42 & --1.70 & --2.16 & --2.15 \\
~~($\lambda$8498)& --2.29 & --1.56 & --1.92 & --1.81 \\
 \ [K/H]       & --2.25\scriptsize{$\pm$0.02} &  --1.65\scriptsize{$\pm$0.03} & --2.17\scriptsize{$\pm$0.04} & --2.08 \\
 \ [Ba/H]      & --3.57 & --2.30 & --3.32 & --3.39 \\
 \ [Sr/H]      & --2.85 & --2.18 & --2.89 & --2.57 \\
 \ [Eu/H]      & --3.16 & --1.61  & & \\
\hline
\end{tabular}
\end{center}
\end{table}

\subsection{\rm HD\,122563}

HD\,122563 is one of the best observed halo stars. The SIMBAD
Astronomical Database finds more than 300 references to studies of
this star since 1980. However, stellar parameters derived by
different authors show a large spread of data. Most papers rely on
an infrared flux method temperature $\Teff$(IRFM) = 4572~K, as
determined by Alonso et al. (\cite{alonso99}). Fulbright
(\cite{fulbright}) gets the lower value, $\Teff$ = 4425~K,
requiring the same iron abundance from the \ion{Fe}{i}\, lines
with high and lower excitation potential.
 A mean temperature $\Teff$ = 4615~K is deduced by Barbuy et al.
(\cite{barbuy}) from using the $b-y$, $V-R$, $V-K$, and $J-K$
calibrations of Alonso et al. (\cite{alonso99}). Surface gravity
derived in different studies, either from the {\sc Hipparcos}
parallax or from the ionization balance between \ion{Fe}{i}\, and
\ion{Fe}{ii}, ranges between $\log g$ = 1.5 (Barbuy et al.
\cite{barbuy}) and $\log g$ = 0.6 (Fulbright \cite{fulbright}).
Significant discrepancies are seen for the microturbulence
velocity: $\Vmic$ varies between 2~\kms\, (Barbuy et al.
\cite{barbuy}) and $\Vmic$ = 2.9~\kms\, (Simmerer et al.
\cite{la2004}).

\begin{figure}
\resizebox{88mm}{!}{\includegraphics{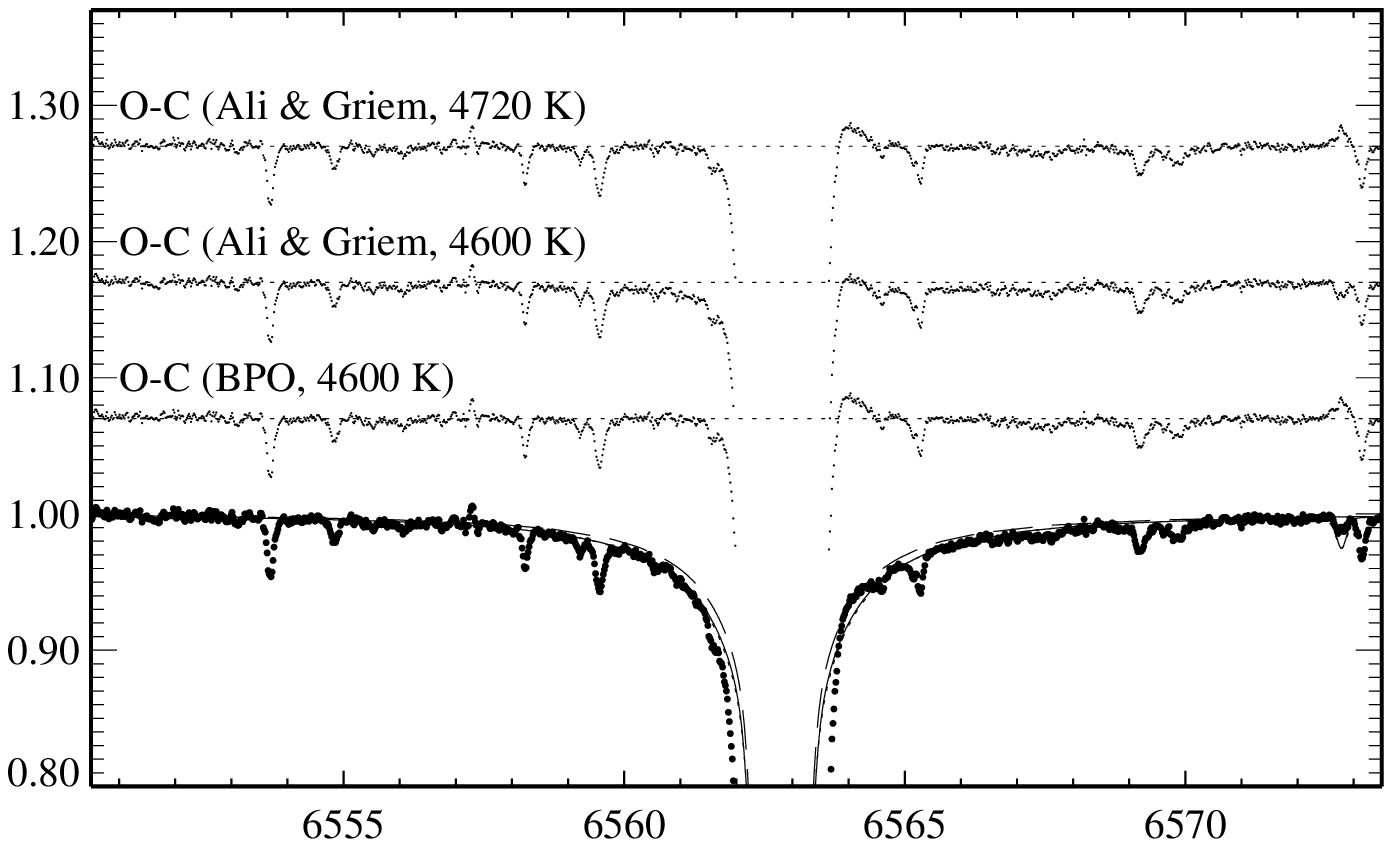}}
\resizebox{88mm}{!}{\includegraphics{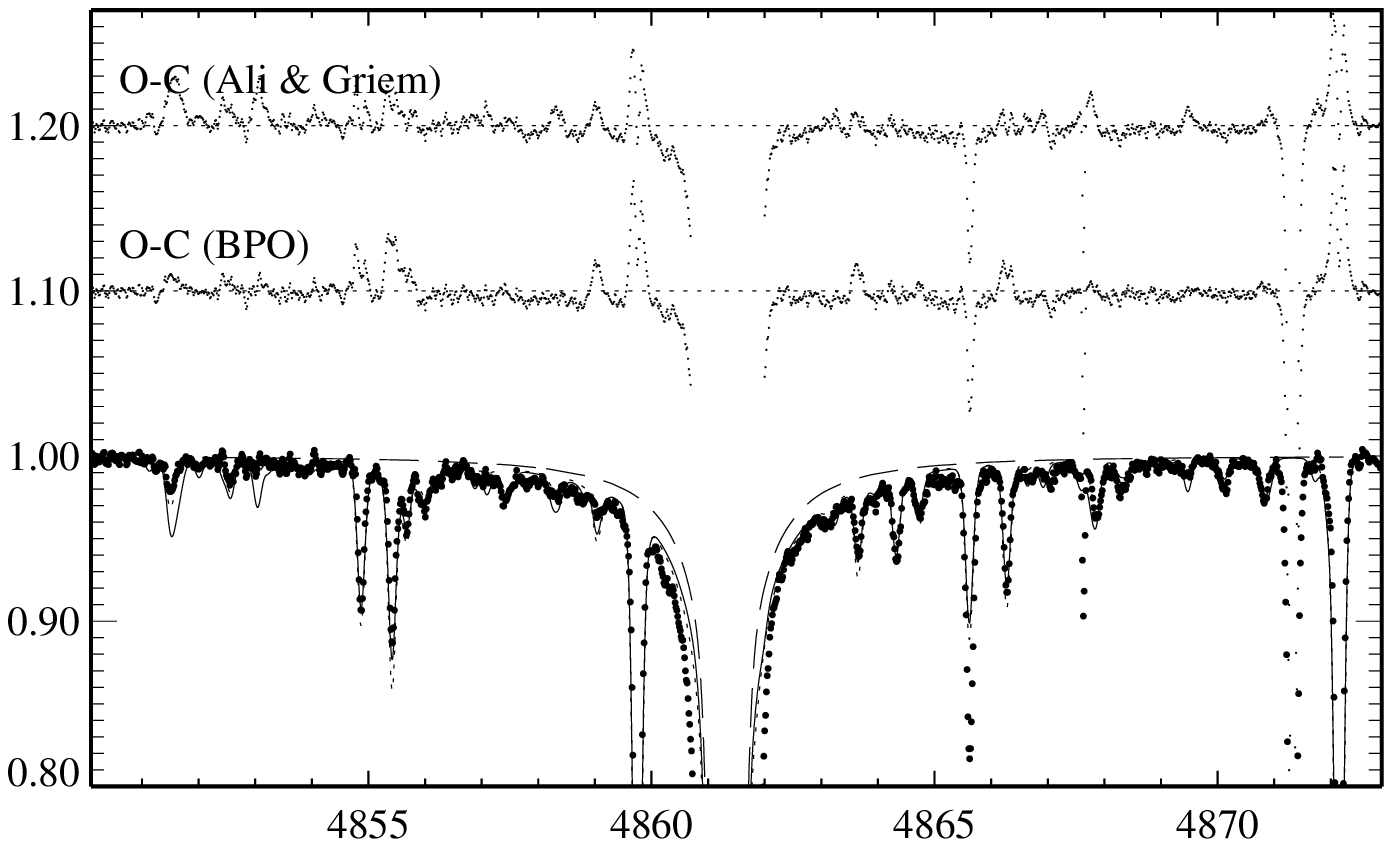}}
\caption[]{ Theoretical non-LTE flux profiles computed using the
$AG$ (continuous line for $\Teff$ = 4720~K, dashed line for $\Teff$ = 4600~K) and $BPO$ (dotted line, $\Teff$ = 4600~K) self-broadening
recipes  compared to the observed UVES profiles (bold
dots) of H$_\alpha$ (top panel) and H$_\beta$ (bottom panel) in
HD\,122563. The (O - C) values are shown in the upper part of each
panel.} \label{h_122563}
\end{figure}

In spectroscopic determination of $\Teff$, we use the observed
spectrum taken from the ESO UVESPOP survey (Bagnulo et al.
\cite{POP03}). The wings of H$_\alpha$ and H$_\beta$ in HD\,122563
are rather weak as compared to the turnoff stars. However, they
are still sensitive to $\Teff$. This is illustrated in
Fig.\,\ref{h_122563}. Non-LTE leads to a consistency of the
temperatures derived from the two Balmer lines. For the two
broadening recipes, we find $\Teff(AG)$ = 4720~K and $\Teff(BPO)$
= 4600~K. The uncertainty of $\Teff$ arising from the profile
fitting is estimated as 60~K for H$_\alpha$ and 50~K for H$_\beta$
\ based on a $S/N$ ratio of the observed spectrum and a
sensitivity of the Balmer lines to a variation of $\Teff$. The
line core within $\pm$1.5~\AA\, for H$_\alpha$ and within
$\pm$1.2~\AA\, for H$_\beta$ is not included in the fit.
Figure\,\ref{h_122563} shows the best fits and the differences
between observed and calculated spectra, (O - C). At the LTE
assumption, H$_\alpha$ gives a 60~K lower temperature.

As a final value we adopt $\Teff$ = 4600~K that is supported both
spectroscopically and photometrically. The uncertainty of 120\,K
is estimated as the difference between $\Teff$ from the two
broadening recipes, because a treatment of the self-broadening of
the hydrogen lines turns out to be the most significant source of
$\Teff$ errors. The effect of $\Teff$ variation by 120~K on metal
abundances is shown in Table\,\ref{uncertain_122563}. We note that
a significant discrepancy of 0.33~dex appears between
\ion{Mg}{i}\, $\lambda4571$ and the remaining five \ion{Mg}{i}\,
lines when $\Teff$ increases by 120~K. This is due to a larger
increase of departures from LTE for the line arising from the
ground state compared to that for the subordinate lines.

\begin{table}[htbp]
\caption{Effect on element abundances (in dex) and the Ba odd
isotope fraction $f_{odd}$ (in \%) of HD\,122563 caused by
uncertainties of its stellar parameters.} \label{uncertain_122563}
\setlength{\tabcolsep}{2.2mm}
\begin{center}
\begin{tabular}{lcccc}
\hline Chemical & $\Teff$(K) & \multicolumn{2}{c}{$\log g$} &
$\Vmic$(\kms) \\
\cline{3-4}
species      & +120  & --0.2 & +0.2   & +0.1 \\
\hline
 \ion{Fe}{ii} & $<+$0.01& --0.07& +0.08 & $<$0.01 \\
 \ion{Ca}{i}  & +0.11 & +0.03 & --0.03 & $<$0.01 \\
 \ion{Ca}{ii} &  &    &    &   \\
 ($\lambda$3933) & +0.11 & +0.02 & --0.01 & $<$0.01 \\
 ($\lambda$8498) & +0.12 & +0.00 & +0.09  & $<$0.01 \\
 \ion{Mg}{i}     & +0.11 & +0.03 & --0.03 & --0.02 \\
 \ion{Ba}{ii}    & +0.07 & --0.08& +0.08  & $<$0.01 \\
 $f_{odd}$           & +12   & +2    & --3    & --8 \\
\hline
\end{tabular}
\end{center}
\end{table}

\begin{figure}
\resizebox{88mm}{!}{\includegraphics{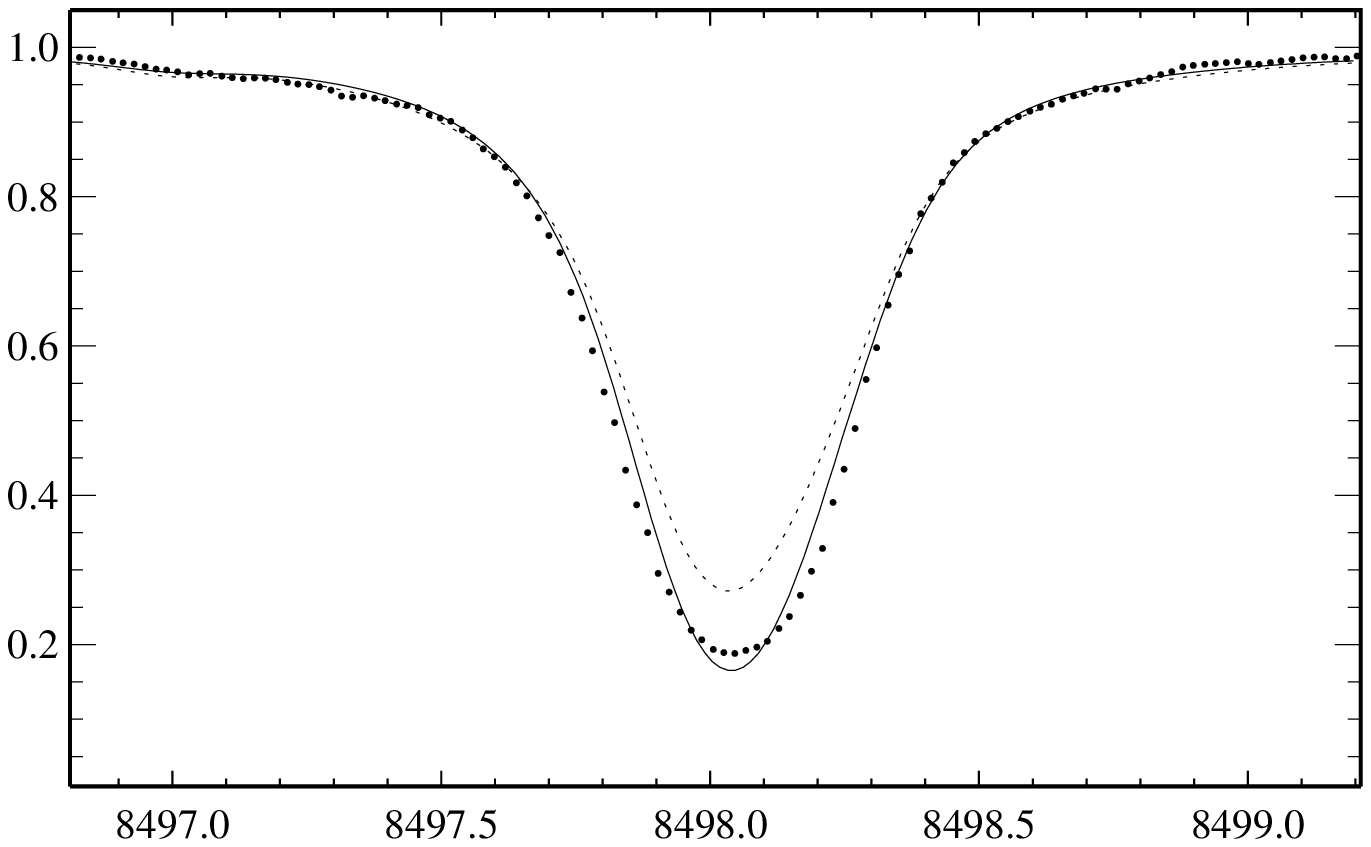}}
\resizebox{88mm}{!}{\includegraphics{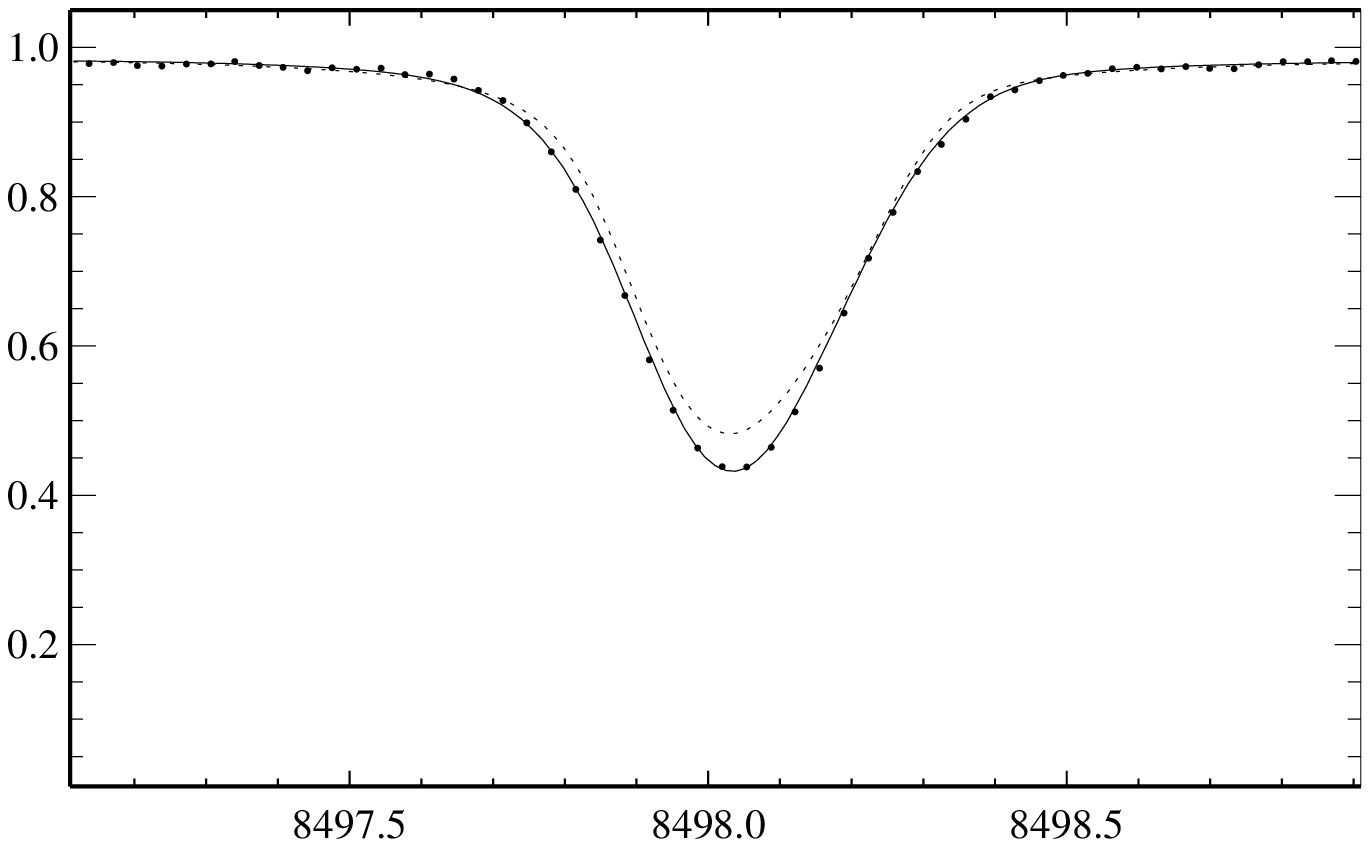}}
\caption[]{The best non-LTE (\kH\ = 0.1, continuous line) and LTE
(dotted line) fits of the observed profiles (bold dots) of
\ion{Ca}{ii}\, $\lambda8498$ in HD\,122563 (top panel) and
BD$+3^\circ$740 (bottom panel). The obtained non-LTE and LTE
abundances of Ca are presented in Table \ref{startab}. Note, that
it is impossible to achieve well fitting in LTE. }
\label{star8498}
\end{figure}

A trigonometric gravity based on the {\sc Hipparcos} parallax,
$\pi = 3.76\pm0.72$~mas, was carefully calculated by Barbuy et al.
(\cite{barbuy}) using $\Teff$ = 4600~K, $\log g = 1.53 \pm 0.2$.
Based on non-LTE line formation, we check the ionization
balance between \ion{Ca}{i}\, and \ion{Ca}{ii}\, for three gravitis,
 $\log g$ = 1.50, 1.30, and 1.70, in order to find a right
value. In each case, we determine the iron abundance from 9 lines,
the Mg abundance from 6 lines, the CaI abundance from 15 lines,
and the microturbulence velocity from the strongest lines of
\ion{Fe}{ii}\, (multiplet 42), \ion{Mg}{i}\, (5 lines), and
\ion{Ca}{i}\, (3 lines). The  \ion{Ca}{ii}\, resonance line is
most probably affected by the chromospheric temperature rise, and
the CaII abundance of HD\,122563 is obtained only from
\ion{Ca}{ii}\, $\lambda8498$ which is believed to be of
photospheric origin. Strong departures from LTE are found for
\ion{Ca}{ii}\, $\lambda8498$, such that, with the LTE assumption,
its profile cannot be fitted by a variation in the Ca abundance.
The LTE abundance is determined from line wing fits as shown in
Fig.\,\ref{star8498}. We note that the theoretical LTE profile in
Fig.\,\ref{star8498} (top panel) has been computed for an 0.07~dex
larger Ca abundance, compared to that for the non-LTE profile, and
this only helps to reproduce the wings but not the core of
\ion{Ca}{ii}\, $\lambda8498$. If to use the equivalent width of
$\lambda8498$, LTE leads to a 0.18~dex overestimated Ca abundance.
In non-LTE, the core of $\lambda8498$ in HD\,122563 is better
fitted though not perfectly. The line is saturated, and its
half-width is only weakly sensitive to a variation of the Ca
abundance and $\Vmic$. We rely therefore on fitting the observed
profile beyond $\Delta\lambda =$0.4\AA. The uncertainty of the
profile fitting translates to the abundance error of 0.06~dex. The
maximal error of the van der Waals damping constant $\Delta\log
C_6$ = 0.2 leads to $\Delta\eps{}$ = 0.06~dex.

The effect of  gravity variation on the abundance of Ca from two
ionization stages is shown in Table\,\ref{uncertain_122563}. We
find that the CaI and CaII abundances are consistent within the
error bars for each of three gravities. The abundance difference
is minimal for $\log g$ = 1.50, (CaI - CaII) = +0.04~dex and
equals +0.07~dex and --0.08~dex for $\log g$ = 1.30 and 1.7,
correspondingly. Thus, a trigonometric gravity is supported
spectroscopically, and we adopt $\log g = 1.50 \pm 0.2$ as a final
value.

\subsection{\rm BD\,$+3^\circ$740}\label{bd3_740}

For BD\,$+3^\circ$740, we note a surprisingly large discrepancy
between the infrared flux method temperatures derived by Alonso et
al. (\cite{alonso96}), $\Teff$(IRFM) = 6110~K, and Mel\'endez \&
Ram\'irez (\cite{melendez04}), $\Teff$(IRFM) = 6440~K. An
effective temperature determined in different studies from
different spectroscopic and photometric methods (for review, see
Ivans et al. \cite{ivans}) varies between $\Teff$ = 6000~K (Ivans
et al. \cite{ivans}) and $\Teff$ = 6355~K (Carretta et al.
\cite{carr02}).

\begin{figure}
\resizebox{88mm}{!}{\includegraphics{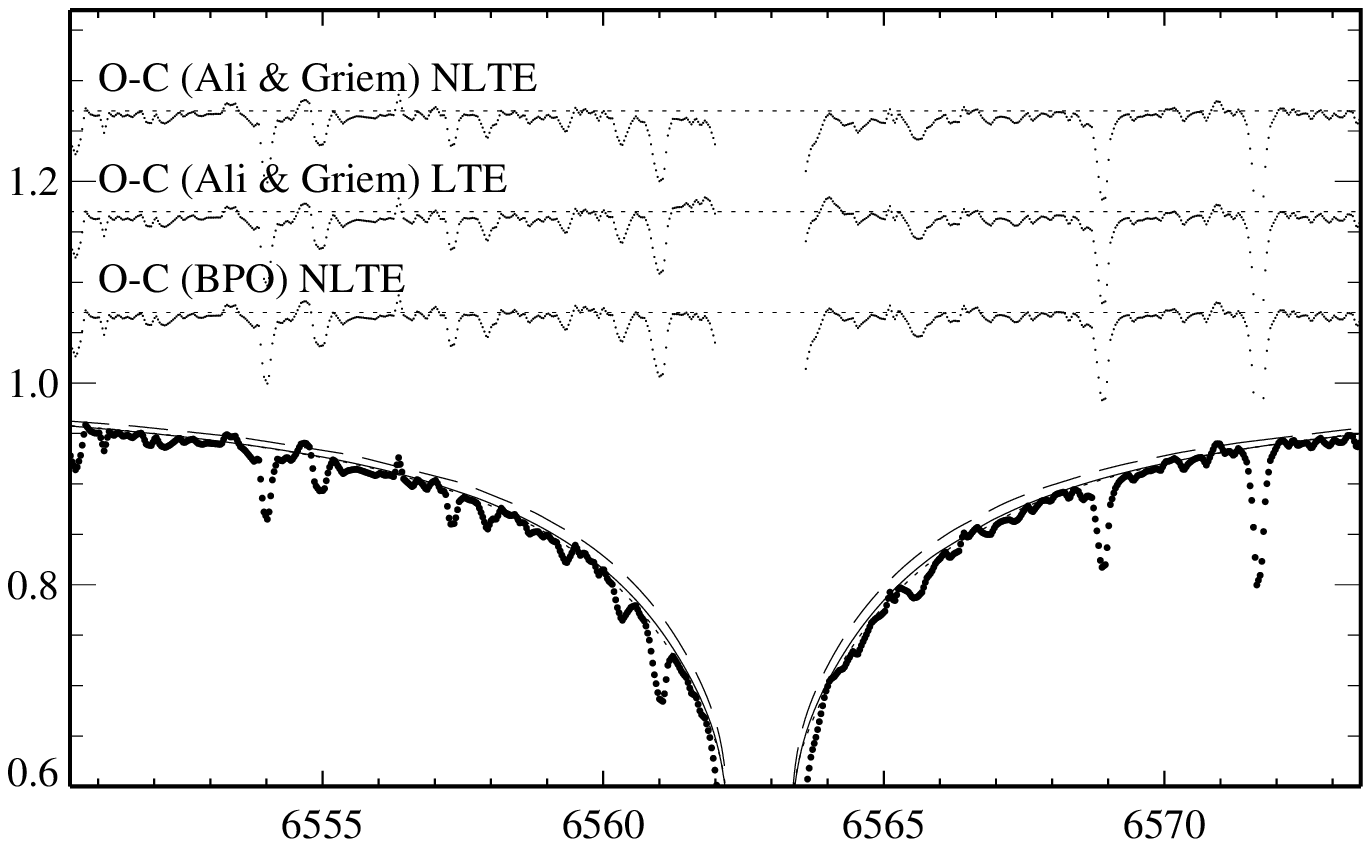}}
\resizebox{88mm}{!}{\includegraphics{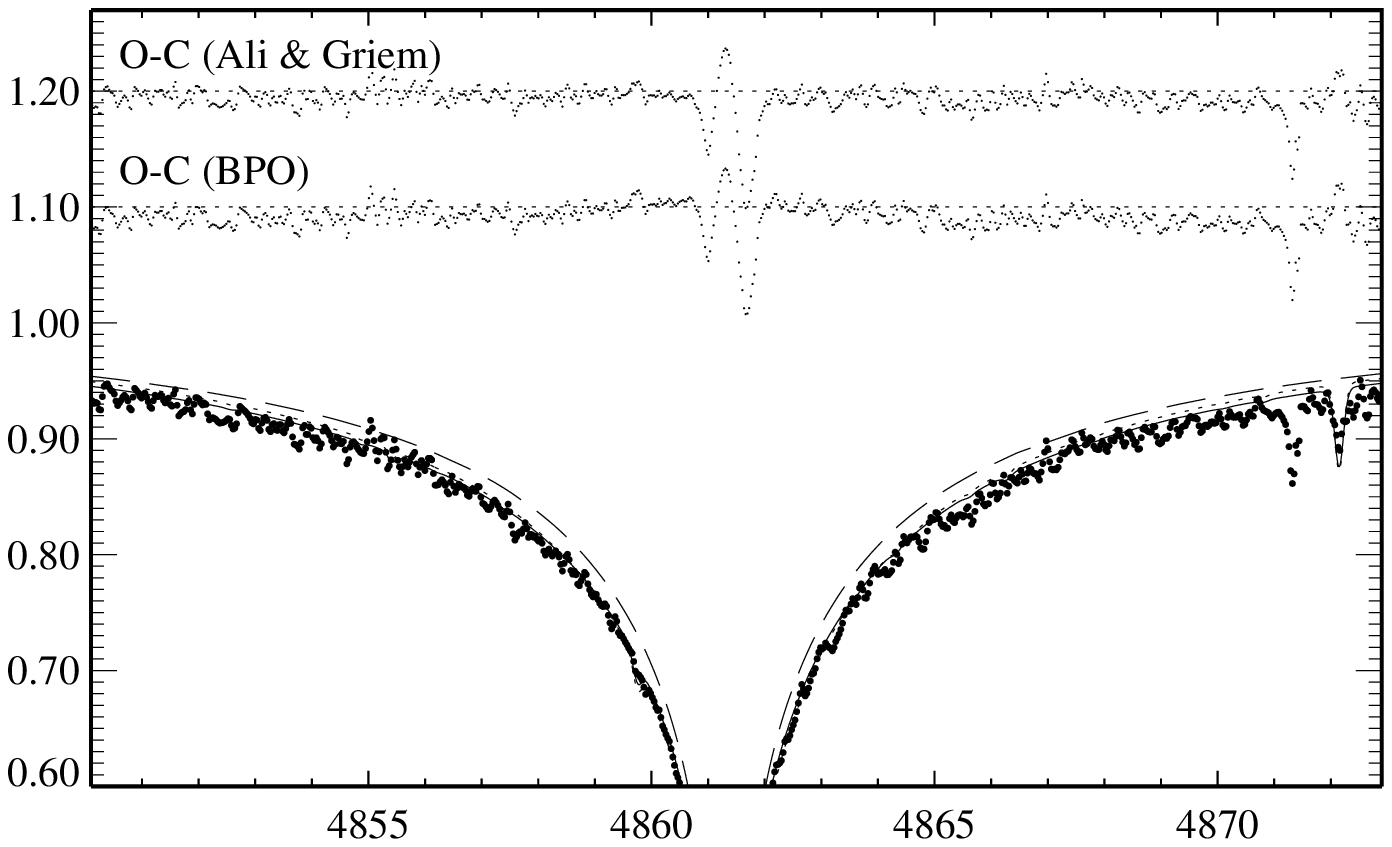}}
\caption[]{ The best non-LTE fits of H$_\alpha$ (top panel) and H$_\beta$ (bottom panel) in
the observed FOCES spectrum (bold dots) of BD$+3^\circ$740 achieved using the
$AG$ (continuous line, $\Teff$ = 6340~K) and $BPO$ (dotted line, $\Teff$ = 6260~K) self-broadening recipes. For comparison, we show (O - C) also for the best LTE $AG$ fit of H$_\alpha$
computed for $\Teff$ = 6280~K. Note that the LTE inner wings ($\Delta\lambda$ = 1.2\AA - 1.6\AA)
are broader compared to the non-LTE ones although the lower
effective temperature is used in the LTE calculations.
 In order to illustrate the sensitivity of the line profiles to $\Teff$, the non-LTE $AG$ profiles are shown also for $\Teff$ = 6240~K (dashed curves). }
\label{h_g84_29}
\end{figure}

Based on non-LTE line formation for \ion{H}{i}\, with hydrogen
collisions taken into account applying \kH\,= 2, we obtain from
both Balmer lines $\Teff(AG)$ = 6340~K and $\Teff(BPO)$ = 6260~K.
The uncertainty of $\Teff$ arising from the profile fitting is
estimated as 70~K for H$_\alpha$ and 80~K for H$_\beta$. The core
within $\pm$1.2~\AA\, for H$_\alpha$ and within $\pm$0.7~\AA\, for
H$_\beta$ is not included in the fit. Figure\,\ref{h_g84_29} shows
the best fits and the sensitivity of the Balmer line profiles to a
100~K variation in $\Teff$. For the LTE calculations, nearly the
same quality fit of H$_\alpha$ is achieved at a 60~K lower
temperature, however, provided that the core within $\pm$1.7~\AA\,
is not included in the fit. In the inner wings ($\Delta\lambda$ =
1.2\AA\, - 1.7\AA), the LTE profile is too broad for $\Teff$
obtained from the far wings (see Fig.\,\ref{h_g84_29}). When
non-LTE calculations are performed with reduced hydrogenic
collisions (\kH\,= 0.1), a 100~K larger effective temperature is
determined from the H$_\alpha$ wings, beyond 2\AA\, from the line
center, while the inner wings are obtained to be too narrow. Thus,
the uncertainty of $\Teff$(H$_\alpha$) due to various treatment of
hydrogen line formation is estimated as (+100~K, --60~K). Having
in mind that non-LTE effects are negligible for H$_\beta$, a
consistency of the temperatures derived from two Balmer lines
favors the non-LTE approach with hydrogen collisions taken into
account with \kH\, of 1 to 2.

As a final value, $\Teff$ = 6340~K, is adopted with the
uncertainty of 100\,K. The effect of $\Teff$ variation on metal
abundances is shown in Table\,\ref{uncertain_g84_29}.

\begin{table}[!t]
\caption{Effect on element abundances (in dex) of
BD\,$+3^\circ$740 and BD\,$-13^\circ$3442 caused by uncertainties
of its stellar parameters} \label{uncertain_g84_29}
\setlength{\tabcolsep}{3.5mm}
\begin{center}
\begin{tabular}{lcccc}
\hline Chemical & $\Teff$(K) & \multicolumn{2}{c}{$\log g$} &
$\Vmic$(\kms) \\
\cline{3-4}
species      & --100  & +0.16 & --0.28  & +0.2 \\
\hline
 \ion{Fe}{ii} & --0.02& +0.05 & --0.09  &$<$0.01 \\
 \ion{Ca}{i}  & --0.06 & +0.02& +0.01   &$<$0.01 \\
 \ion{Ca}{ii} &  &    &       \\
 ($\lambda$3933) & --0.10 & --0.03 & +0.04 &$<$0.01 \\
 ($\lambda$8498) & --0.08 & +0.06 & --0.05 & --0.06 \\
\hline
\end{tabular}
\end{center}
\end{table}

According to Fuhrmann (\cite{Fuhr98}), a trigonometric gravity
corresponding to the {\sc Hipparcos} parallax, $\pi =
7.80\pm2.09$~mas, is inferred as $\log g$ = 4.18, and it can lie
between $\log g$ = 3.92 and $\log g$ = 4.37 due to the relatively
large error of the parallax. Asplund et al. (\cite{asplund06}) use
the mean value of the absolute visual magnitude derived from the
{\sc Hipparcos} parallax and from Str\"omgren photometry and
obtain $\log g$ = 4.04. In both studies, $\Teff$ = 6260~K is used.
We have checked the ionization balance between \ion{Ca}{i}\, and
\ion{Ca}{ii}\, for $\log g$ = 4.06 and 3.90, applying $\Teff$ =
6340~K. Table\,\ref{uncertain_g84_29} presents the effect of
gravity varying on the obtained abundance of Ca. The CaI abundance
is determined from 10 lines with the measurement error $\sigma$ =
0.05~dex. For \ion{Ca}{ii} $\lambda8498$, $\Delta\Vmic$ =
0.2~\kms\, leads to the abundance error of 0.06~dex, while the
uncertainty of the profile fitting to 0.03~dex. For the
\ion{Ca}{ii} resonance line, the total abundance error is
estimated as 0.06~dex.

For $\log g$ = 3.90, we find a consistency within 0.04~dex of the
CaI and CaII abundances. For the higher gravity, an agreement
between the different lines of Ca is destroyed: CaII
($\lambda8498$) - CaI = 0.09~dex and CaII ($\lambda8498$ ) - CaII
($\lambda3933$) = 0.11~dex. As a final value, we adopt $\log g$ =
3.90$\pm$0.15 which is the lower limit of a trigonometric gravity
and which is best supported spectroscopically.

\subsection{\rm BD\,$-13^\circ$3442}

This is the faintest star of our small sample with no {\sc
Hipparcos} parallax measured. In the observed UVES spectrum of
this star, only H$\alpha$ can be used to derive a temperature. The obtained spectroscopic temperature
based either on the Ali \& Griem (\cite{AG66}) theory, $\Teff(AG)$
= 6390~K, or on the $BPO$ recipe, $\Teff(BPO)$ = 6310~K,
 is lower compared to an
infrared flux method temperature, $\Teff$(IRFM) = 6484~K, as
determined by Mel\'endez \& Ram\'irez (\cite{melendez04}). As a
final value, $\Teff$ = 6390~K is adopted with the uncertainty similar to that for BD\,$+3^\circ$740, 100~K.

Using the absolute visual magnitude from Str\"omgren photometry,
Asplund et al. (\cite{asplund06}) obtain $\log g$ = 3.86 for
BD\,$-13^\circ$3442. A 0.02~dex larger surface gravity is obtained
for a slightly higher temperature, $\Teff$ = 6390~K, determined in
this study. Assuming that the gravity error is similar to that
found for BD\,$+3^\circ$740, we check the ionization balance
between \ion{Ca}{i}\, and \ion{Ca}{ii}\, for $\log g$ = 3.88 and
3.60. For each of two gravities, the CaI and CaII abundances turn
out to be consistent within the error bars. The effect of a
gravity variation on the obtained abundances of Ca is shown in
Table\,\ref{uncertain_g84_29}. We adopt $\log g$ = 3.88 as a final
value. It is worth noting that the calculated atmospheric
parameters of BD\,$-13^\circ$3442 and BD\,$+3^\circ$740 are very
similar.

It is interesting to inspect the position of the selected stars on
evolutionary tracks using the obtained stellar parameters. The
grid of stellar models of VandenBerg et al. (\cite{isohrone}) is
used. According to interpolation along lines of constant radii in
a cube of masses, metal abundances and [$\alpha$/Fe] ratios for
available tracks we derive the HRD position of HD\,84937 with a
mass of 0.784 $M_\odot$ and an age of 14.0~Gyr (linear
interpolation) or an age of 13.7~Gyr (logarithmic interpolation).
The age determined for the known trigonometric parallax is
slightly higher, i.e. 15.3~Gyr. The other two turnoff stars,
BD$+3^\circ 740$ and BD$-13^\circ 3442$, are outside the set of
evolutionary tracks with their low metal abundances. Their masses
and ages can only be estimated by extrapolation (the lowest metal
abundance of the grid is at [Fe/H] = -2.3). Their masses are 0.766
$M_\odot$ (BD$+3^\circ 740$) and 0.790 $M_\odot$ (BD$-13^\circ
3442$), whereas their ages are 14.9 Gyr (BD$+3^\circ 740$) and
13.0 Gyr (BD$-13^\circ 3442$). The estimated ages are acceptable
in view of the WMAP results that center around an age of the
universe of 13.7~Gyr (Spergel et al. \cite{spergel2003}).
HD\,122563 lies on the giant branch. In such case, extrapolation
leads to too uncertain results for stellar ages.

\section{Abundance analysis}\label{Abundance}

In order to derive aluminium and potassium abundances, we use the
UVES archive spectra because the \ion{Al}{i}\, $\lambda3961$ and
\ion{K}{i}\, $\lambda\lambda$7665, 7699 lines are not covered by
the available Subaru spectra. For BD\,$+3^\circ$740, we cannot
calculate Al abundance from the line at $\lambda3961$\AA\ because
the observed spectrum in this region is missing. The \ion{K}{i}\,
$\lambda7699$ line in BD$-13^\circ$3442 is blended with telluric
lines and therefore is excluded from analysis. In
BD\,$+3^\circ$740 and BD$-13^\circ$3442, the \ion{Eu}{ii}\,
$\lambda4129$ and \ion{Ba}{ii}\, $\lambda6496$ lines cannot be
extracted from noise. The Ba abundance of these stars is obtained
from a single line, \ion{Ba}{ii}\, $\lambda4554$, assuming a pure
$r-$process even-to-odd Ba isotope ratio of 54 : 46 (Arlandini et
al. \cite{rs99}). It is worth noting that the \ion{Ba}{ii}\, line
in these stars is weak ($W_\lambda \sim 10$~m\AA) and is
insensitive to a variation of Ba isotopic abundances. The derived
non-LTE and LTE abundances are presented in Table\,\ref{startab}.

For the selected elements, non-LTE abundances of the two stars
were determined in our earlier studies based on the FOCES observed
spectra (Gehren et al. \cite{gehren06}, Na and Mg in HD\,84937 and
BD$+3^\circ$740; Mashonkina et al. \cite{mash_ca}, Mg and Ca in
HD\,84937). Shimanskaya \& Mashonkina (\cite{shim_mg}) calculated
the non-LTE abundances of Mg for HD\,84937 and
BD$+3^\circ$740 based on the equivalent widths from the
literature. In this paper, we obtain very similar results.

 We find that HD\,122563 reveals significantly lower element/Fe
abundance ratios compared to the remaining stars. For the most
investigated elements, there is underabundance by 1.2~dex for Eu, by more than 0.75~dex
for Ba, 0.5~dex for Al, 0.3~dex for K and Sr, and 0.2~dex for Ca.
This reflects most probably a chemical inhomogeneity of the
interstellar medium in the
early Galaxy. An environment, where HD\,122563 was formed, was
enriched with the neutron-capture elements and the elements with
odd nuclear charge Al and K to the less extent compared to the
interstellar gas out of which the remaining stars were formed.

\subsection{Departures from LTE}

Significant non-LTE effects are found for almost all lines in our
program stars. Non-LTE abundance corrections $\Delta_{\rm NLTE} =
\eps{non-LTE}-\eps{LTE}$ are negative for alkali atoms \ion{Na}{i} and \ion{K}{i}. They grow
in absolute value with the strength of the observed line, i.e.
with decreasing effective temperature. The maximal values are
reached for HD\,122653: $\Delta_{\rm NLTE}$(\ion{Na}{i}) =
--0.44~dex and $\Delta_{\rm NLTE}$(\ion{K}{i}) = --0.35~dex.

\begin{figure}
\resizebox{88mm}{!}{\includegraphics{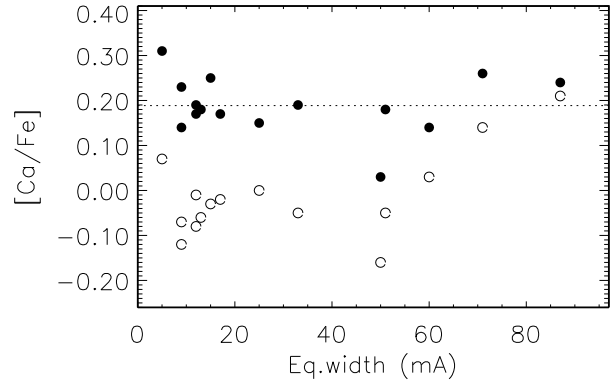}}
\caption[]{ Trends of non-LTE (filled circles) and LTE (open
circles) abundance with line strength determined from the
\ion{Ca}{i}\, lines in HD\,122563. Note the steep trend of the LTE
[Ca/Fe] values with line strength above $W_\lambda$ = 60\,m\AA.
The mean non-LTE value is shown by dotted line.} \label{ca1_122563}
\end{figure}

For the photoionization dominated minority species \ion{Mg}{i},
\ion{Al}{i}, and \ion{Ca}{i}, $\Delta_{\rm NLTE}$ is positive. We
note a strong temperature dependence of departures from LTE for
\ion{Al}{i}. Non-LTE abundance correction is equal 0.86~dex for
the hottest star BD$-13^\circ$3442, while consists of only
0.04~dex for the coolest star HD\,122653. For the \ion{Mg}{i}\,
and \ion{Ca}{i}\, lines, $\Delta_{\rm NLTE}$ does not exceed
0.2~dex. However, an advantage of the non-LTE approach is clearly
seen in Fig.\,\ref{ca1_122563} where we show the Ca abundances
calculated from the neutral Ca lines in HD\,122653. Non-LTE
removes the steep trend of the [Ca/Fe] values with line strength
obtained under the LTE assumption above $W_\lambda$ = 60\,m\AA.

As was found theoretically in our earlier studies, non-LTE may
lead either to strengthening or to weakening the \ion{Ba}{ii}\,
(Mashonkina et al. \cite{Mash99}) and \ion{Sr}{ii}\, (Mashonkina
\& Gehren \cite{eubasr}) lines depending on stellar parameters and
element abundance. This is nicely illustrated with our small
sample of stars. A slightly negative non-LTE abundance correction
is obtained for \ion{Sr}{ii}\, $\lambda4215$ in HD\,122653 and a
positive one for two the most metal-poor stars, while $\Delta_{\rm
NLTE}$ = 0 for HD\,84937. For the \ion{Ba}{ii}\, lines in our
program stars, $\Delta_{\rm NLTE}$ is positive and reaches
0.50~dex for BD$-13^\circ$3442. Non-LTE effects lead to $\Delta_{\rm NLTE}$ =
+0.16~dex for \ion{Eu}{ii}\, $\lambda4129$.

\subsection{Comparisons with non-LTE studies}

For the elements under unvestigation, systematic analyses based on
non-LTE line formation are few in number. We find that the non-LTE
abundances obtained in this paper agree within the error bars with
the corresponding abundances in the literature for common stars.
For the three stars, HD\,84937, BD$+3^\circ$740, and
BD$-13^\circ$3442, the absolute abundance difference (this study -
Idiart \& Thevenin, \cite{th-mg}) does not exceed 0.08~dex for Mg
and 0.12~dex for Ca. The non-LTE abundances of Mg and Ca given by
Gratton et al. (\cite{gr_mg}) for HD\,84937 are consistent within
0.06~dex with ours.

The non-LTE abundance of Na determined by Mishenina et al.
(\cite{mish_na}) for HD\,84937 agrees within the error bars with
our data. For HD\,122563, we derive a 0.14~dex larger Na abundance
compared to that given by Andrievsky et al. (\cite{Andr}). This is
most probably due to the lower surface gravity ($\log g$ = 1.1)
used in their study and, therefore, the stronger non-LTE effects
for the \ion{Na}{i} lines.

The non-LTE abundance of potassium obtained by
Takeda et al. (\cite{Ta-k}) agrees within 0.03~dex with our value for HD\,122563.

The non-LTE abundances of Al and heavy elements Sr, Ba, and Eu are
determined for the first time for the program stars.

\subsection{Galactic abundance trends: non-LTE vs. LTE}

In the view of significant departures from LTE found for the
program stars, it is important to investigate what might change
from the abundance trends determined from the larger samples of
stars analyzed in LTE in the literature.

{\it The $\alpha-$process elements}, Mg and Ca. There are well
established observational evidences for that, in old metal-poor
([Fe/H] $< -1$) stars of the Galaxy, Mg and Ca are overabundant
relative to iron (see e.g. recent determinations of Fulbright
\cite{fulbright}, Cayrel et al. \cite{cayr04}, Cohen et al.
\cite{cohen2004}, Barklem et al. \cite{bark05}, they all computed
under LTE). The different samples of stars reveal close together
average abundance ratios for Mg/Fe and Ca/Fe and the small scatter
of data at the level of the measurement errors. E.g. Cohen et al.
(\cite{cohen2004}) obtain the mean abundances [Mg/Fe] =
0.39$\pm$0.13 and [Ca/Fe] = 0.31$\pm$0.12 for the sample of the
turnoff stars with $-3.5 \le$ [Fe/H] $\le -2$, and Cayrel et al.
(\cite{cayr04}) find [Mg/Fe] = 0.27$\pm$0.13 and [Ca/Fe] =
0.33$\pm$0.11 for the sample of the cool giants with $-4 \le$
[Fe/H] $\le -2.4$. In the range of stellar parameters the studies
are concerned with, non-LTE leads to larger abundances of Mg and
Ca derived from the lines of neutral atoms. According to Zhao \&
Gehren (\cite{mgzh01}, \ion{Mg}{i}), Idiart \& Thevenin
(\cite{th-mg}, \ion{Mg}{i} and \ion{Ca}{i}), and Mashonkina et al.
(\cite{mash_ca}, \ion{Ca}{i}), non-LTE abundance corrections may
reach 0.2~dex depending on stellar parameters. For the Cohen et
al. sample, they are expected to be similar to that calculated for
our program turnoff stars BD$+3^\circ$740 and BD$-13^\circ$3442,
$\Delta_{\rm NLTE}$ = +0.04~dex for Mg and $\Delta_{\rm NLTE}$ =
+0.14~dex for Ca. Based on our calculations for HD\,122563, we
expect that the LTE abundances of Mg and Ca in the cool giants of
the Johnson (\cite{johnson2002}) and Cayrel et al. (\cite{cayr04})
samples may be larger in non-LTE by approximately 0.2~dex. We note
that accounting for departures from LTE leads to nearly identical
overabundances of Mg and Ca relative to Fe in the turnoff stars of
Cohen et al. and the cool giants of Cayrel et al. with the Ca/Mg
abundance ratio close to solar value. This makes the conclusions
of the cited papers about a similarly constant ratio between the
yields of iron and the $\alpha-$process elements Mg and Ca in the
early Galaxy more solid.

{\it Aluminium.} As for neutral magnesium and calcium, non-LTE
leads to larger abundances of Al derived from the resonance lines
of \ion{Al}{i} compared to those from LTE analysis. However,
departures from LTE for \ion{Al}{i} depend strongly on effective
temperature and metallicity. As was shown above, the
metal-deficient turnoff stars are affected most strongly by Al
abundance corrections reaching +0.86~dex for BD$-13^\circ$3442.
When non-LTE effects are neglected, Galactic abundance trends have
a large scatter of data for the stars of close metallicity. E.g.
for their sample of VMP stars ($-3.8 \le$ [Fe/H] $\le -1.5$),
Barklem et al. (\cite{bark05}) find the scatter of the order of
0.5~dex for the [Al/Fe] and [Al/Mg] ratios, while the scatters are
less than the typical abundance errors for [Mg/Fe] and [Ca/Fe].
The data of Fulbright (\cite{fulbright02}) on [Al/Fe] reveals even
larger scatter up to 1~dex at [Fe/H] $< -2$.

Some abundance analyses (e.g. Carretta et al. \cite{carr02},
Cayrel et al. \cite{cayr04}, Cohen et al. \cite{cohen2004})
carried out under the LTE assumption use the non-LTE abundance
corrections published in Baum\"uller \& Gehren (\cite{al_corr}).
We note that Baum\"uller \& Gehren (\cite{al_corr}) give
$\Delta_{\rm NLTE}$ for the small grid of model atmospheres, and
individual corrections obtained by interpolating for given stellar
parameters should be considered as a coarse treatment of the
non-LTE effects. Nevertheless even a rough approach leads to a
dramatical change of Galactic abundance trends. For the sample of
stars in the metallicity range $-3.6 \le$ [Fe/H] $\le -2.4$,
Carretta et al. obtain the average LTE abundance [Al/Fe] =
--0.59$\pm$0.15, while the Al abundance increases up to [Al/Fe] =
+0.10$\pm$0.22 when the non-LTE corrections are applied. A
different case is the Cayrel et al. study. They use an
inappropriate correction for their sample of cool giants,
$\Delta_{\rm NLTE}$ = +0.65~dex that corresponds to the model with
$\Teff$ = 5500~K, $\log g$ = 3.5, [Fe/H] = --3. Our calculations
for HD\,122563 show that the Al abundance correction is much
smaller, $\Delta_{\rm NLTE}$ = +0.04~dex. The non-LTE abundances
of Al presented by Cayrel et al. in their Fig.~8 are therefore
overestimated.

{\it Alkalis}, \ion{Na}{i} and \ion{K}{i}. Contrary to Mg, Ca, and
Al, non-LTE leads to smaller abundances of sodium and potassium
compared to those determined in LTE. As was shown by Baum\"uller
et al. (\cite{nlte_na}) and Andrievsky et al. (\cite{Andr}) and
supported in this paper, departures from LTE for \ion{Na}{i}
depend strongly on the surface gravity and the element abundance
of a star. For the cool giant HD\,122563, the non-LTE abundance
correction of \ion{Na}{i} $\lambda5889$ / $\lambda5895$ is a
0.15~dex / 0.20~dex more negative compared to that for the turnoff
star BD$-13^\circ$3442 of close metallicity and Na abundance. For
our program turnoff stars, $\Delta_{\rm NLTE}$ consists of
--0.47~dex / --0.39~dex for the atmospheric parameters that match
HD\,84937 and decreases in absolute value with decreasing
metallicity, by approximately 0.2~dex for $\Delta$[Fe/H] = --0.5.
As for \ion{Na}{i}, the non-LTE effects for \ion{K}{i} increase
with decreasing surface gravity. For alkalis, a full treatment of
non-LTE line formation is therefore required for each individual
star. Andrievsky et al. (\cite{Andr}) show a clear advantage of
such an approach compared to the use of a constant non-LTE
abundance correction. They achieve the small star-to-star
variation in the [Na/Fe] and [Na/Mg] ratios comparable with the
measurement errors and consistent results for the samples of VMP
dwarfs and unmixed giants.

Stellar abundance ratios between the odd-Z and even-Z elements
such as Na/Mg, Al/Mg, and K/Ca are important for testing current
nucleosynthesis models. We emphasize that exactly these abundance
ratios are most affected by departures from LTE. E.g. non-LTE
leads to an increase of the Al/Mg ratio in the program turnoff
stars by 0.77~dex. In contrary, the Na/Mg ratio decreases, on
average, by 0.42~dex. It is worth noting that our non-LTE
abundance ratios Al/Mg and Na/Mg agree well with the Galaxy
chemical evolution calculations of Timmes et al. (\cite{timmes})
for the corresponding metallicity range. An exception is a too low
value of [Al/Mg] in HD\,122563. As was noted, this star reveals an
overall deficiency of the elements requiring for their synthesis a
large flux of neutrons.

Our data on potassium with the mean non-LTE ratio [K/Ca] =
--0.19$\pm$0.06 favor the Galaxy chemical evolution calculations
of Goswami \& Prantzos (\cite{goswami}). Similar conclusion was
drawn by Shimansky et al. (\cite{shiman}) and Zhang et al.
(\cite{nlte_kb}) based on non-LTE determinations of the potassium
abundance in the mildly metal-poor ([Fe/H] $> -1$) stars.

Due to the strong influence of non-LTE effects in the atmospheres
of metal-deficient stars LTE abundance analyses of Na, Al, and K
are nearly useless for an investigation of Galactic chemical
evolution.

{\it Neutron-capture elements} \ Sr, Ba, and Eu. The LTE abundance
analyses (see e.g. Fulbright \cite{fulbright02}, Carretta et al.
\cite{carr02}, Barklem et al. \cite{bark05}) determine a steep
decline of the [Ba/Fe] abundance ratios at [Fe/H] $< -2$. Based on
our non-LTE calculations we expect that the trend might be
shallower in non-LTE. E.g. for the program turnoff stars, [Ba/Fe]
decreases from 0.00 at [Fe/H] = --2.15 down to --0.23 at [Fe/H] =
--2.66 in non-LTE, while from --0.15 down to --0.73 in LTE.

As found in many studies (for a review of the literature under
2004, see Travaglio et al. \cite{travaglio04} and the later study
of Barklem et al. \cite{bark05}, they all computed under LTE), the
abundance trends of Sr/Fe, Ba/Fe, Eu/Fe, and the ratios among
neutron-capture elements have a large scatter at fixed metallicity
that appears to increase with decreasing [Fe/H]. A part of the
dispersions may be due to ignoring deviations from LTE in
abundance determinations, because the non-LTE effects for the
lines of \ion{Sr}{ii} and \ion{Ba}{ii} depend strongly on stellar
parameters and the element abundance.

\begin{figure}
\resizebox{88mm}{!}{\includegraphics{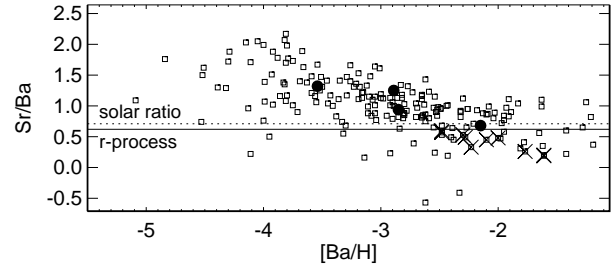}}
\caption[]{A run of Sr/Ba vs. [Ba/H] for the stars of Barklem et
al. (\cite{bark05}) sample (open squares) and our program stars
(filled circles). Here, Sr/Ba  = $\log(N_{Sr}/N_{Ba})$ where $N_X$ are number densities. The $r-$process rich (r-II) stars are marked by
crosses. The solar total abundance ratio Sr/Ba = 0.71 obtained in this study is indicated by dotted line and the relative solar system $r-$process abundance
(Sr/Ba)$_r = 0.62$ according to Arlandini et al. (\cite{rs99}) by
solid line.} \label{srba}
\end{figure}

The measured scatter is usually interpreted as cosmic in origin,
because iron and neutron-capture elements are produced in
different sites. We suggest to plot the ratios among
neutron-capture elements against the abundance of Ba, but not the
Fe abundance. In Fig.~\ref{srba}, we combine the Sr/Ba abundance
ratios from Barklem et al. (\cite{bark05}) with our data. Except
for the few outliers, the scatter at fixed Ba abundance does not
exceed 0.5~dex at [Ba/H] $> -3.5$ and approaches to 1~dex at the
lower [Ba/H]. This makes possible to see a growth of Sr/Ba towards
the lower Ba abundance. For comparison, the scatter in the run
[Sr/Ba] - [Fe/H] (see Fig.~22 from Barklem et al.) approaches to
1.5~dex at [Fe/H] $< -2.5$ and hides the correlations between the
abundance ratios. According to the recent study of Mashonkina et
al. (\cite{sr_y_zr}), the Zr/Ba and Y/Ba abundance ratios in halo
stars grow towards the lower Ba abundance. A clear abundance trend
seen in Figure~\ref{srba} indicates an existence of the process
that yielded in the early Galaxy strontium with little
contribution to Ba. The present study supports the ideas of Truran
et al. (\cite{truran02}) and observational findings of Aoki et al.
(\cite{aoki05}) and  Honda et al. (\cite{honda}) that the light
and heavy neutron-capture elements in VMP stars come from distinct
processes. Synthesis of the light neutron-capture elements remains
a challenge to theorists.

\section{The r/s-process controversy for HD\,84937 and HD\,122653}

\subsection{[Eu/Ba] abundance ratios}

It is well known (among many observational studies see e.g. Spite
\& Spite \cite{Spite} and Barklem et al. \cite{bark05}) that in
old metal-poor galactic stars europium is overabundant relative to
Ba. This is interpreted as a a dominance of the $r-$process in
production of heavy elements beyond the iron group in the early
Galaxy. The solar system abundance ratio of Eu to Ba contributed
by the r-process relative to the total abundances, [Eu/Ba]$_r$,
equals $0.70$ according to Arlandini et al. (\cite{rs99}). In
several studies, Sneden et al. (\cite{sned96, sned00}), Cowan et
al. (\cite{Cowan99, Cowan02}), and Hill et al. (\cite{hill}) have
presented arguments supporting constant relative r-process element
abundances during the history of the Galaxy (at least, where $56
\le Z < 70$). In two stars, BD$+3^\circ$740 and BD$-13^\circ$3442,
the Eu abundance is not determined. Our computations for [Eu/Fe] =
0.5 (which translates to [Eu/Ba] = 0.7) predict the central depth
of \ion{Eu}{ii}\, $\lambda4129$ in these stars at a level of
0.5\%. Such a weak line cannot be measured in the spectra
available to us. Thus, the Eu abundance in BD$+3^\circ$740 and
BD$-13^\circ$3442 may be large enough to give a [Eu/Ba] ratio
close to 0.7. We obtain that HD\,84937 reveals abundances close to
a pure $r-$process production of heavy elements with [Eu/Ba] =
0.70. For HD\,122563, the corresponding abundance ratio is
smaller, [Eu/Ba] = 0.50, that suggests a contribution from the
main $s-$process to observed barium.

\subsection{The fraction of the odd isotopes of Ba}\label{isotope}

A determination of the Ba even-to-odd isotope abundance ratio in
stars becomes possible due to the significant hyperfine structure
(HFS) affecting the \ion{Ba}{ii}\, resonance lines of the odd
isotopes. In total, the resonance line $\lambda4554$ has 15
components spread over 58 m\AA. The larger the fraction of the odd
isotopes, the stronger the HFS broadening of $\lambda4554$ and the
larger the energy absorbed in this line. The HFS broadening of the
\ion{Ba}{ii}\, subordinate lines is small. The maximal effect on
the derived abundance is found for $\lambda6496$ and it does not
exceed 0.01~dex for the Sun and is much smaller for our program
metal-poor stars. We use the method described in detail in our
previous study (Mashonkina \& Zhao \cite{ba_iso}) and find the Ba
even-to-odd isotope abundance ratio in a given star from the
requirement that Ba abundances derived from the \ion{Ba}{ii}\,
resonance line and the subordinate lines must be equal.

For both stars, HD\,84937 and HD\,122563, the only subordinate
line, \ion{Ba}{ii}\, $\lambda6496$, is used to determine total Ba
abundance. It is given in Table\,\ref{startab}. The \ion{Ba}{ii}\,
$\lambda5853$ line cannot be extracted from noise in HD\,84937 and
is affected by a hot pixel in HD\,122563. We do not use
\ion{Ba}{ii}\, $\lambda6141$ due to blending with a \ion{Fe}{i}\,
line.

Having fixed the Ba abundance of HD\,84937 ([Ba/Fe] = 0) and
varying the even-to-odd isotope abundance ratio, we obtain a
consistent element abundance from $\lambda4554$ for a fraction of
the odd isotopes $f_{odd} = 0.43\pm0.14$. The uncertainty of the
desired value is mainly caused by the uncertainty of the
microturbulence velocity, which is estimated as +0.2\,\kms. Recent
$r-$process calculations of Kratz et al. (\cite{kratz}) predict
$f_{odd} = 0.44$. A close value, $f_{odd} = 0.46$, is obtained
from $r-$residuals in the stellar model of Arlandini et al.
(\cite{rs99}). Thus, both the [Eu/Ba] abundance ratio and the
even-to-odd Ba isotope abundance ratio suggest a pure $r-$process
production of heavy neutron-capture elements in HD\,84937.

For HD\,122563, we find $f_{odd} = 0.22\pm0.15$ with the
uncertainty mainly caused by the uncertainties of $\Teff$ (120~K)
and $\Vmic$ (0.1\,\kms) (see Table\,\ref{uncertain_122563}). This
fraction is close to the solar system value (0.18). Taking into
account the 1$\sigma$ error, we estimate the $r-$process
contribution to Ba in HD\,122563 to be no larger than 70\% in
agreement with the [Eu/Ba] abundance ratio in this star. The only
halo star with a measured fraction of the odd isotopes of Ba is
HD\,140283 (Lambert \& Allende Prieto \cite{iso140283}). Similarly
to HD\,122563, it shows a lower value $f_{odd} = 0.31$ compared
with that of two other halo stars with a measured fraction of the
odd isotopes of Ba, HD\,84937 and HD\,103095 ($f_{odd} = 0.42$,
Mashonkina \& Zhao \cite{ba_iso}). We note that both HD\,140283
and HD\,122563 reveal lower neutron-capture element abundances
compared to those of the stars of close metallicity and look as
$r-$process-poor stars. For HD\,140283, [Ba/Fe] = --0.80 according
to Mashonkina et al. (\cite{sr_y_zr}). Using the same stellar
parameters $\Teff$ = 5810~K, $\log g$ = 3.68, and [Fe/H] = --2.43,
we estimate [Eu/Fe] = --0.31 based on the UVESPOP observed
spectrum (Bagnulo et al. \cite{POP03}) and non-LTE line formation.
The obtained ratio [Eu/Ba] = 0.49 is smaller than the value
typical of the halo stars (e.g. [Eu/Ba]  = 0.69 deduced by
McWilliam \cite{mcw98} as the mean ratio for the sample of stars
with [Fe/H] $\leq -2.4$) and smaller than the relative solar
system $r-$process abundance [Eu/Ba]$_r$ = 0.70 (Arlandini et al.
\cite{rs99}).

\section{Conclusions}\label{conclusion}

In this paper, we determine for the first time stellar parameters
$\Teff$ and $\log g$ of four VMP stars ($-2.15 \le$ [Fe/H] $\le
-2.66$) based on non-LTE line formation. Effective temperature is
derived from the Balmer H$_\alpha$ and H$_\beta$ line wing fits.
For each star, non-LTE leads to a consistency of $\Teff$ from two
lines and to a higher temperature compared to the LTE case, by up
to 60~K.

We find a clear advantage of the non-LTE approach in spectroscopic
determination of surface gravity compared to the LTE case. For
each of three stars with {\sc Hipparcos} parallax available, a
surface gravity obtained from the non-LTE ionization balance
between \ion{Ca}{i} and \ion{Ca}{ii} agrees within the error bars
with a trigonometric one. The \ion{Ca}{i}/\ion{Ca}{ii} line
intensity ratio involving \ion{Ca}{ii} $\lambda8498$ is
particularly sensitive to a variation of surface gravity. However,
the wings of the \ion{Ca}{ii} resonance lines can also be used for
spectroscopic determination of $\log g$ of VMP stars. We conclude
that non-LTE calculations based on ${\rm 1D}$ model atmospheres
provide reliable results for $\Teff$ and $\log g$ of VMP stars in
the wide range of temperatures ($\Teff$ = 4600~K to 6400~K) and
gravities ($\log g$ = 4.0 down to 1.5).

Due to strong influence of non-LTE effects in the atmospheres of
VMP stars abundances of Na, Mg, Al, K, Ca, Sr, Ba, and Eu cannot
be determined in a reliable way assuming LTE. Departures from LTE
depend strongly on stellar parameters such that a precise
abundance analysis is only possible on the base of non-LTE
calculations for each individual star.

We derive the fraction of the odd isotopes of Ba in two stars,
HD\,84937 and HD\,122563, and, thus, increase the number of halo
stars with a measured $f_{odd}$ to four. Two stars among four have
low $f_{odd}$: HD\,122563 with $f_{odd}$ = 0.22$\pm$0.15 (this
study) and HD\,140283 with $f_{odd}$ = 0.31 (Lambert \& Allende
Prieto \cite{iso140283}). Both stars reveal the lower Eu/Ba ratios
compared to the relative solar system $r-$process abundances.
These results suggest that the $s-$process has significantly
contributed to the Ba abundance in these stars, or that the
unknown process that produced high Sr/Ba ratios in these stars
yielded the low $f_{odd}$ value. Two remaining stars have the
larger fraction of the odd isotopes of Ba: $f_{odd}$ =
0.43$\pm$0.14 in HD\,84937 (this study) and $f_{odd}$ = 0.42 in
HD\,103095 (Mashonkina \& Zhao \cite{ba_iso}). In both stars, the
Eu/Ba ratios are consistent with the relative solar system
$r-$process abundances. We conclude therefore, that $f_{odd}$
found in these stars can serve as observational constraint on
$r-$process models.

\begin{acknowledgements}
The authors thank Manuel Bautista for providing with the new
collisional data on \ion{Ca}{ii} and Keith Butler for his help.
 ML acknowledges with gratitude
the National Astronomical Observatories of Chinese Academy of
Science for warm hospitality during a productive stay in September
of 2006. This research was supported by the Russian Foundation for
Basic Research with grant 05-02-39005-GFEN-a, the Natural Science
Foundation of China with grants NSFC 10433010, 10521001, and
10778612, the Deutsche Forschungsgemeinschaft with grant 436 RUS
17/13/07, the RF President with a grant on Leading Scientific
Schools 784.2006.2, and the Presidium RAS Programme ``Origin and
evolution of stars and galaxies''.
We thank the anonymous referee for constructive suggestions and useful remarks.
\end{acknowledgements}

\end{document}